\documentclass{article}
\PassOptionsToPackage{numbers, compress}{natbib}

\usepackage[preprint]{neurips_2026}

\usepackage[utf8]{inputenc} 
\usepackage[T1]{fontenc}    
\usepackage[breaklinks=true,  colorlinks,  bookmarks=false]{hyperref}
\usepackage{url}            
\usepackage{booktabs}       
\usepackage{enumitem}       
\usepackage{amsfonts}       
\usepackage{nicefrac}       
\usepackage{microtype}      
\usepackage{xcolor}         
\usepackage{graphicx}
\usepackage{amsmath}
\usepackage{listings}
\title{STAR-PólyaMath: Multi-Agent Reasoning \\ under Persistent Meta-Strategic Supervision}

\author{%
  \textbf{Jiaao Wu}$^{1}$\thanks{Work done during internship at Microsoft Research.} \quad
  \textbf{Xian Zhang}$^{2}$\footnotemark[2] \quad
  \textbf{Hanzhang Liu}$^{3}$ \\
  \textbf{Sophia Zhang}$^{4}$ \quad
  \textbf{Fan Yang}$^{2}$ \quad
  \textbf{Yinpeng Dong}$^{1}$\thanks{Corresponding authors.} \\[0.6ex]
  $^{1}$Tsinghua University \quad
  $^{2}$Microsoft Research \quad
  $^{3}$New York University \quad
  $^{4}$MIT \\[0.4ex]
  \texttt{wuja25@mails.tsinghua.edu.cn} \quad
  \texttt{dongyinpeng@mail.tsinghua.edu.cn} \\
  \texttt{\{zhxian,~fanyang\}@microsoft.com} \quad
  \texttt{hl4963@nyu.edu} \quad
  \texttt{zsophia@mit.edu} \\
}

\begin{document}

\maketitle

\begin{abstract}
  Frontier AI models and multi-agent systems have led to significant improvements in mathematical reasoning. However, for problems requiring extended, long-horizon reasoning, existing systems continue to suffer from fundamental reliability issues: hallucination accumulation, memory fragmentation, and imbalanced reasoning-tool trade-offs. In this paper, we introduce \textbf{STAR-PólyaMath}, a multi-agent framework that systematically addresses these challenges through meta-level supervision and structured Reasoner-Verifier interaction. STAR-PólyaMath is structured as an orchestrated state machine with nested challenge-step-replan loops, governed by a reasoning-free Python orchestrator that separates control from inference and bounds error propagation through trace-back and re-planning. Our key innovation is a \textbf{persistent Meta-Strategist} that maintains cross-attempt memory and exercises meta-level control by issuing high-level strategic guidance or mandatory directives, so the system can escape unproductive loops rather than stagnate or over-rely on tools. STAR-PólyaMath achieves state-of-the-art results on all eight top-tier competition benchmarks: AIME 2025-2026, MathArena Apex Shortlist, MathArena Apex 2025, Putnam 2025, IMO 2025, HMMT February 2026, and USAMO 2026. It obtains perfect scores on AIMEs, Putnam, and HMMT, and shows its largest margin on Apex 2025, scoring 93.75\% compared with 80.21\% by the strongest baseline GPT-5.5. Ablation studies show that the gains arise from the framework's orchestration rather than from model-level diversity since removing key components or substituting in mixed backbones consistently weakens performance. Code is available at \url{https://github.com/Julius-Woo/STAR-PolyaMath}.
\end{abstract}

\section{Introduction}

Large Language Models (LLMs) now perform strongly on competition mathematics, with frontier systems reaching near-perfect scores on Olympiad-style benchmarks and delivering strong results on the Putnam and FrontierMath~\citep{openai2025, google2025, dekoninck2025, glazer2024}. Progress has come from two complementary directions: natural-language reasoning strengthened by chain-of-thought prompting and reinforcement learning with verifiable rewards~\citep{wei2022cot, deepseekr1}, and tool-augmented reasoning via theorem provers or executable code~\citep{moura2021, alphaproof, seedprover2025, gou2024tora, gao2023pal}. Nevertheless, long-horizon problems remain challenging because free-form reasoning is difficult to verify, while tool use often incurs substantial computational and temporal costs.

A central limitation lies in the difficulty of reliable self-correction: LLMs often fail to revise flawed reasoning without external feedback, and explicit self-correction can even degrade performance~\citep{huang2024selfcorrection}. This limitation has motivated multi-agent frameworks that decompose reasoning into specialized roles, typically pairing a solution generator with a verifier that provides critique signals~\citep{lei2024macm}. Such verification-centered workflows, including training-free verify-and-refine pipelines~\citep{imo2025gold} and multi-agent debate~\citep{khan2024debate}, have demonstrated strong performance on competition-level mathematics. Related agentic systems further incorporate tool use~\citep{autogen2023}, enabling code execution or formal verification during problem solving. Collectively, recent training-free and training-based agentic frameworks~\citep{hilbert2025, seedprover2025, toolorchestra2025}, together with frontier proprietary models~\citep{openai2025, google2025}, suggest that multi-agent systems have become a leading paradigm for complex mathematical reasoning.

Despite these advances, existing systems continue to exhibit three recurring failure modes:
\begin{itemize}[leftmargin=*]\vspace{-1ex}
  \item \textbf{Hallucination accumulation.} Errors in intermediate reasoning propagate across long solution trajectories, as models tend to generate confident but incorrect arguments. Although debate~\citep{khan2024debate} and self-verification~\citep{deepseekmath2025} can mitigate this effect, reliably discriminating superficially plausible reasoning from ultimately invalid derivations remains difficult~\citep{mahdavi2025brains}.\vspace{-1ex}
  \item \textbf{Memory fragmentation.} Difficult problems often require repeated backtracking across multiple unsuccessful lines of attack, yet most agentic systems either retain excessive context or fail to preserve salient information about prior attempts. Although certain formal systems address this through lemma caching and recursive decomposition~\citep{seedprover2025, hilbert2025}, these solutions are specialized for formal verification and do not generalize to open-ended mathematical reasoning.\vspace{-1ex}
  \item \textbf{Imbalanced reasoning-tool trade-offs.} Although code execution is reliable, excessive reliance on brute-force search can obscure mathematical structure and lead to intractable exploration. Prior work shows that models trained on tool-use trajectories develop a systematic bias toward code~\citep{wang2025tocode}. Existing remedies typically require additional training or ensemble overhead~\citep{toolorchestra2025, drori2025diverse}. In practice, agents often lack \emph{metacognition}, the ability to step back and revise their strategy, leading to inefficient computation and failures on abstract problems.\vspace{-1ex}
\end{itemize}

Frontier agentic AI models have already possessed the knowledge and capacity to solve long-horizon mathematical competition problems. We argue that the missing ingredient is \emph{persistent meta-level supervision}. In human problem-solving, a strong supervisor does not carry out every technical step, but guides at a meta level: selecting promising proof strategies, tracking failed approaches, and intervening when progress stalls. We instantiate this role as a \textbf{persistent Meta-Strategist} that retains global oversight throughout the entire problem-solving process. It maintains strategic memory across attempts and issues \emph{meta-strategies}: decisions about the problem-solving strategy, formed as either high-level guidance grounded in domain knowledge or mandatory directives when the agents enter unproductive loops, such as repeatedly generating code instead of actively reasoning.

\begin{figure}[tb]
\centering\vspace{-2ex}
\includegraphics[width=0.95\linewidth, trim=5pt 1pt 5pt 5pt, clip]{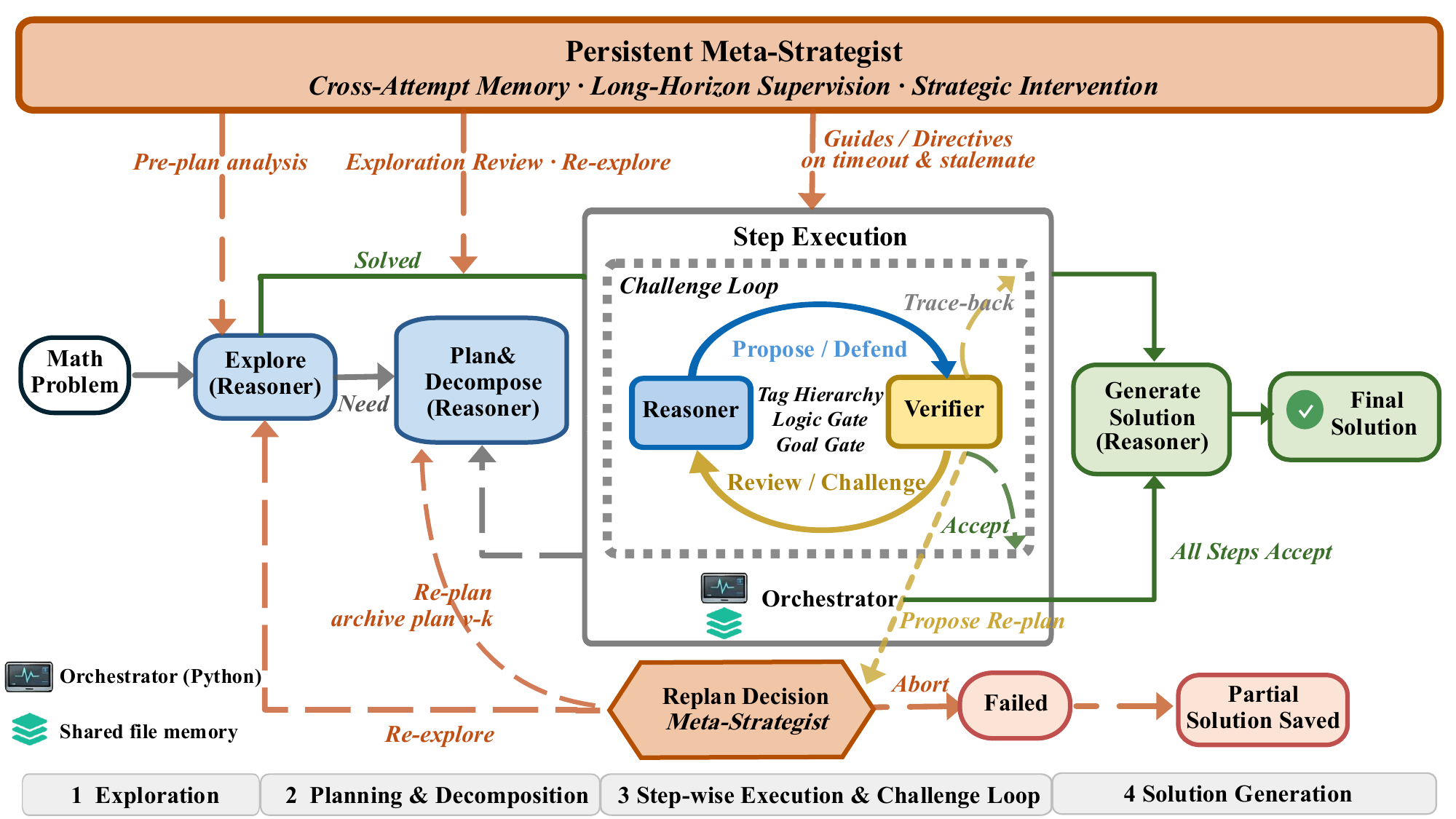}\vspace{-1ex}
\caption{\textbf{STAR-PólyaMath system workflow.} STAR-PólyaMath advances each problem with four phases: \emph{exploration}, \emph{planning and decomposition}, \emph{step-wise execution with challenge loops}, and \emph{solution generation}. A Python orchestrator dispatches the LLM agents and decides advance, trace-back, re-plan, and abort transitions. The Reasoner probes the problem, proposes a step-wise plan, and executes each step with hierarchical verification tags (\texttt{[verified]} / \texttt{[easy-verify]} / \texttt{[hard-verify]}). The Verifier reviews via logic and goal gates, and emits \textsc{Accept}, \textsc{Challenge}, \textsc{Trace-Back}, or \textsc{Propose-Replan}; persistent Reasoner-Verifier debate resolves disagreements within a step, while trace-back and re-planning bound error propagation across steps. A persistent Meta-Strategist supervises the trajectory with cross-attempt memory, issuing \emph{pre-planning analysis}, \emph{strategic guidance or mandatory directives} on timeout or stalemate, and \emph{re-plan} verdicts.}\vspace{-2ex}
\label{fig:architecture}
\end{figure}

Based on this design, we introduce \textbf{STAR-PólyaMath}\footnote{The name pays homage to George Pólya's \emph{How to Solve It}~\citep{polya1945}, which decomposes problem solving into \emph{understanding the problem}, \emph{devising a plan}, \emph{carrying out the plan}, and \emph{looking back}, and emphasizes heuristics such as working on analogous or auxiliary problems and revising the plan when it stops making progress. Our \emph{exploration}, \emph{planning}, \emph{step-wise execution with verification}, and \emph{re-plan} phases instantiate the first three; the persistent Meta-Strategist enacts the ``look back'' stage by examining the trajectory across attempts and switching to an auxiliary plan when the current one stalls.}, an agentic framework with three LLM roles: a \emph{Reasoner}, a \emph{Verifier}, and a persistent \emph{Meta-Strategist}, all coordinated by a reasoning-free Python orchestrator, as illustrated in Figure~\ref{fig:architecture}. The orchestrator controls execution flow: it advances a problem through an explicit state machine with nested challenge-step-replan loops, dispatching the Reasoner and Verifier through structured bidirectional debate loops, while the Meta-Strategist supervises the entire trajectory and intervenes through strategic guidance or mandatory directives. Empirically, this combination yields consistent gains across both final-answer and proof-based benchmarks: STAR-PólyaMath achieves the best reported scores across AIME 2025, MathArena Apex Shortlist, MathArena Apex 2025, Putnam 2025, IMO 2025, AIME 2026, HMMT February 2026, and USAMO 2026, with its largest margin on MathArena Apex 2025 (93.75 vs.\ 80.21 by the strongest baseline GPT-5.5). 
We further conduct systematic ablation studies to demonstrate that each component of the framework contributes meaningfully to the overall performance, and that swapping the standard shared backbone for mixed-model configurations does not improve results, indicating the gains from the framework's orchestration rather than from model selection.

\vspace{-0.5ex}
\section{Related Work}
\vspace{-0.5ex}

\textbf{Mathematical Reasoning with LLMs.} Early neural theorem proving focused on formal systems such as Lean and Isabelle \citep{moura2021}. Recent work has shifted toward informal reasoning: chain-of-thought prompting~\citep{wei2022cot} and self-consistency~\citep{wang2023selfconsistency} improve multi-step problem solving, while program-aided methods such as PAL \citep{gao2023pal} and ToRA \citep{gou2024tora} integrate code execution. Open-source models have also reached IMO gold-medal performance: DeepSeekMath-V2 \citep{deepseekmath2025} uses Group Relative Policy Optimization (GRPO) with meta-verification to internalize generation and verification in one model. Code-based Self-Verification (CSV) \citep{zhou2024solving} grounds probabilistic LLM outputs via deterministic Python execution, bridging informal reasoning and formal verification. In contrast, our work introduces a multi-agent architecture with specialized roles and persistent memory.

\textbf{Multi-Agent Systems for Reasoning.} Debate-based approaches show that model argumentation improves accuracy \citep{du2024debate,khan2024debate}. CAMEL~\citep{li2023camel} and AutoGen~\citep{autogen2023} explore role-playing agents for collaborative problem-solving. MACM~\citep{lei2024macm} introduces condition mining with multi-agent coordination. Recent training-based approaches include MALT~\citep{motwani2024malt}, which trains a Generator-Verifier-Refiner pipeline through multi-agent reinforcement learning, and ReMA~\citep{wan2025rema}, which uses hierarchical meta-reasoning with a high-level planner and low-level solver. However, these systems typically embed control inside one of the reasoning agents and reset context between sessions, leaving no component that supervises across attempts. STAR-PólyaMath's contribution is a structured separation of concerns: a reasoning-free Python orchestrator owns control flow, the Reasoner and Verifier interact through a structured debate protocol, and a persistent Meta-Strategist maintains cross-attempt memory and exercises meta-level supervision over the whole trajectory.

\textbf{Self-Correction and Verification.} Self-refinement \citep{madaan2023selfrefine} explores iterative self-improvement, but self-verification has fundamental limits: the model that erred often cannot detect its own errors~\citep{huang2024selfcorrection}. Recent work on LLM evaluation \citep{mahdavi2025brains,petrov2025proof} reveals that models frequently fail to distinguish correct reasoning from plausible-sounding but flawed arguments, underscoring verification challenges. STAR-PólyaMath addresses this through structured Reasoner-Verifier debate with full session continuity and explicit verification tags that scale scrutiny to claim type, so that code-grounded results, easily checkable claims, and purely mathematical arguments are each evaluated appropriately.

\textbf{Competition-Level Mathematics.} AlphaGeometry~\citep{trinh2024alphageometry} combines neural models with symbolic deduction for geometry olympiads, achieving silver-medal performance at IMO. AlphaProof~\citep{alphaproof} uses reinforcement learning with Lean for formal theorem proving. FunSearch~\citep{romera2024funsearch} employs LLM evolutionary search for mathematical discovery. Agentic reasoning frameworks~\citep{wu2025agentic} integrate tool use with multi-agent coordination. Analyses of code-reasoning trade-offs~\citep{wang2025tocode} reveal systematic biases toward brute-force approaches. STAR-PólyaMath differs by targeting all competition mathematics (not just geometry) and using debate-based verification with persistent metacognitive oversight.

\vspace{-1ex}
\section{Method}
\vspace{-1ex}

In this section, we present the design of STAR-PólyaMath. We first present the system architecture in Section~\ref{sec:2-1}, then provide detailed descriptions of the problem-solving workflow, including \emph{exploration} (Section~\ref{sec:2-explore}), \emph{planning and decomposition} (Section~\ref{sec:2-2}), \emph{step-wise execution with the challenge loop} (Section~\ref{sec:2-3}), and \emph{solution generation} (Section~\ref{sec:2-4}).

\subsection{System Overview}\label{sec:2-1}

STAR-PólyaMath is implemented on top of a CLI agent platform that natively supports code execution, file editing, and session resumption. The system has three specialized agents coordinated by a Python orchestrator. This separation between control and reasoning enables robust long-horizon mathematical reasoning. Implementation details are given in Appendix~\ref{app:impl}.

\textbf{Orchestrator (Python).} Owns all control flow and governs agent interactions. It maintains the per-problem state (current phase, plan, accepted results, prior failures), advances the problem through an explicit state machine, dispatches the LLM agents at the appropriate transitions, and parses their outputs into deterministic \textsc{Advance} / \textsc{Trace-Back} / \textsc{Re-Plan} / \textsc{Abort} decisions.

\textbf{Reasoner.} The primary problem-solving agent. It explores the problem to gain insights, proposes a step-wise plan, executes each step (in natural language, symbolic manipulation, or executable code), defends its arguments in debate, and writes the final solution. It is not assumed to be reliable in isolation and interacts closely with the Verifier through a structured challenge loop.

\textbf{Verifier.} A strict reviewer evaluating the correctness and consistency of the Reasoner's outputs. For each step, it issues one of four verdicts: \textsc{Accept} to proceed, \textsc{Challenge} with specific objections, \textsc{Trace-Back to step $M$} to re-examine earlier reasoning, or \textsc{Propose-Replan} when no choice of step output could rescue the current plan. To mitigate verifier brittleness, the Reasoner is allowed to respond to challenges through a structured debate protocol with full session continuity before any control decision is committed.

\textbf{Meta-Strategist.} A human-like supervisor that operates at the meta level of the reasoning process. Unlike the Reasoner and Verifier, whose sessions can be reset or re-instantiated during execution, the Meta-Strategist maintains a single persistent session for the entire problem lifecycle, so that prior attempts, abandoned strategies, and chronic failure patterns remain in its working context. It issues \emph{meta-strategies} at key decision points about which strategy the agents should pursue, retire, or revisit, either as light strategic guidance or, when agents enter unproductive loops, as mandatory directives. It is also the sole authority on re-planning verdicts: any proposal to abandon the current plan is routed to it for a specific decision.

\subsection{Exploration}\label{sec:2-explore}

Before committing to a fresh new problem-solving plan, STAR-PólyaMath enters an \emph{exploration} phase in which the Reasoner spends a bounded budget on cheap, non-committal investigation, such as enumerating small cases, looking for patterns, drafting candidate conjectures, and recording approaches that immediately fail. Each round produces a deliverable with a self-assigned assessment \textsc{Solved}, \textsc{Partially-Solved}, or \textsc{Need-Plan} which the orchestrator uses to decide what comes next.

If the Reasoner self-reports \textsc{Solved}, the orchestrator immediately routes the polished sketch through a whole-solution review by the Verifier; if accepted, the problem terminates here without ever entering the planning phase. In practice, this approach allows relatively simple problems to be solved quickly, improving overall system efficiency. \textsc{Partially-Solved} rounds are reviewed by the Meta-Strategist, who decides whether to invest in another exploration round or to proceed to plan with the accumulated evidence. \textsc{Need-Plan} and \textsc{Unknown} rounds fall through directly to planning. Across all branches, the structured findings of every round (what was tried, what worked, what failed, candidate approaches) are persisted as shared context that subsequent phases consume.

This phase is also re-entered after a re-plan. When the Meta-Strategist authorizes abandoning a plan, the orchestrator can run a fresh, evidence-driven exploration round before a new plan is generated, so that the next plan does not silently re-anchor on the same failed conjecture as the previous one.

\subsection{Planning and Decomposition}\label{sec:2-2}

Given the problem and any accumulated exploration findings, planning optionally begins with a \emph{pre-planning analysis} in which the Meta-Strategist selects applicable methods from a predefined strategy library (e.g., invariant-based reasoning), flags common pitfalls, and warns of likely misinterpretations. These suggestions enter the Reasoner's context.

The Reasoner then formulates an explicit plan by decomposing the problem into a sequence of numbered steps (typically 6--10), where each step corresponds to a sub-goal. The orchestrator validates only the plan’s structure, checking that the steps are properly numbered, parseable, and arranged in a coherent sequence, without evaluating the mathematical content. If the plan is structurally invalid, the Reasoner is asked to revise it; otherwise the system proceeds to step-wise execution.

This explicit decomposition mitigates \emph{hallucination accumulation} by forcing the solution process to proceed through smaller, inspectable sub-goals rather than a single unconstrained reasoning trace.

\subsection{Step Execution and Challenge Loop}\label{sec:2-3}

STAR-PólyaMath executes the plan in a step-wise manner. For each step, the Reasoner produces a detailed report containing the proposed reasoning, computations, or constructions. Each report is annotated with a \emph{verification tag} indicating the appropriate level of scrutiny:
\begin{itemize}[leftmargin=*]\vspace{-1ex}
    \item \texttt{[verified]}: results obtained from executed Python code and treated as directly grounded;\vspace{-1ex}
    \item \texttt{[easy-verify]}: claims that can be checked with code snippets or straightforward calculations;\vspace{-1ex}
    \item \texttt{[hard-verify]}: claims that require careful mathematical review.\vspace{-1ex}
\end{itemize}
The use of hierarchical verification tags supports a principled trade-off between computation and reasoning: claims suited for reliable execution are grounded in code, while purely mathematical arguments are subjected to stricter logical scrutiny.

The Verifier reviews the report through a \emph{two-gate} evaluation. The \emph{Goal Gate} compares the step's stated goal against what the report actually delivered, guarding against \emph{semantic drift} where locally valid arguments miss the step (e.g., proving only one direction of a sharpness claim). The \emph{Logic Gate} then checks correctness on the report's own terms, scaling scrutiny to each verification tag. Failure of either gate cannot be accepted. The Verifier emits one of four verdicts: \textsc{Accept}, \textsc{Challenge}, \textsc{Trace-Back to step $M$}, or \textsc{Propose-Replan}. The orchestrator dispatches based on the verdict; the four mechanisms below cover all non-\textsc{Accept} cases.

\textbf{Challenge Loop.} When a step is challenged, the Reasoner and Verifier enter a structured debate with full session continuity. The Reasoner may defend, clarify, or revise its argument; the Verifier re-evaluates and re-issues a verdict each round. Both agents retain the in-step debate history so the same argument is not litigated twice. The loop terminates either when the Verifier accepts the revised report or when a fixed maximum number of rounds is reached without convergence; the latter is treated as a \emph{stalemate} signal and surfaced to the Meta-Strategist.

\textbf{Trace-Back.} When the Verifier determines that an error originates from an earlier step $M$, the orchestrator archives steps $M, M{+}1, \ldots$ along with their debate logs, resets the Reasoner session so the next attempt does not carry abandoned context forward, and re-enters the step loop at step $M$. Verified intermediate results are auto-rescued and made available to the new attempt.

\textbf{Re-Plan.} A re-plan can be \emph{proposed} by any of three sources: the Verifier \textsc{Propose-Replan}, the Reasoner emits explicit \texttt{[plan-blocked]} tag in a response, or the orchestrator (a stalemate, chronic step-execution timeouts, or a Meta-Strategist intervention that already requested replanning). All proposals are routed to a single \emph{Replan Decision} made by the Meta-Strategist, who returns one of four verdicts: \textsc{Continue}, \textsc{Trace-Back to step $M$}, \textsc{Approve-Replan} (archive the current plan, mark previously failed directions forbidden, and start a new plan attempt), or \textsc{Abort} (terminate the problem). This unified verdict mechanism prevents the system from re-planning into directions it has already tried and abandoned, and keeps the Meta-Strategist as the single point of authority on plan-level decisions.

\textbf{Timeouts and pure-reasoning mode.} When a long-running computation exceeds its time budget, the Meta-Strategist intervenes by extracting partial results and providing strategic guidance for the next attempt. A particularly forceful directive it can issue is to switch the Reasoner into a \emph{pure-reasoning mode} that disables further code execution, forcing it to analyze the existing computational evidence by mathematical means. This mechanism directly addresses the \emph{reasoning-tool trade-off} by enabling metacognitive intervention precisely when brute-force tool use becomes counterproductive.

Together, these mechanisms reduce \emph{hallucination accumulation} by enforcing early error detection, preventing downstream error propagation through trace-back, and enabling the system to escape invalid lines of reasoning via re-planning. They also mitigate \emph{memory fragmentation}: the challenge loop preserves debate context within each step, file-based progress tracking maintains continuity across session resets, and the Meta-Strategist's persistent session lets it recognize recurring failure patterns across attempts and respond with history-informed interventions rather than generic guidance.

\subsection{Solution Generation}\label{sec:2-4}

Once all steps have been accepted, the Reasoner consolidates the accepted step reports into a complete final solution. The Meta-Strategist can perform an optional \emph{whole-solution review}, evaluating the assembled solution as a single document. This guard catches problems that step-level review can miss, such as cross-step inconsistencies. If the whole-solution review rejects the draft, the orchestrator re-invokes the Reasoner with the rejection visible up to a small retry budget, and otherwise saves the solution and terminates with success.

\section{Experiments}

We evaluate STAR-PólyaMath on competition-level mathematical reasoning benchmarks, including recently released 2026 contests. These benchmarks represent the pinnacle of mathematical competition, featuring the most recent and challenging problems that serve as rigorous tests of AI capabilities.

\subsection{Setup}\label{sec:setup}

\textbf{Benchmarks.} Five benchmarks are \emph{final-answer}: \textbf{AIME 2025} and \textbf{AIME 2026} (30 problems each from the American Invitational Mathematics Examination, integer answers in $000$--$999$); \textbf{HMMT February 2026} (33 problems from the Harvard--MIT Mathematics Tournament); and \textbf{MathArena Apex 2025} (12 problems) together with \textbf{Apex Shortlist} (48 problems)~\citep{matharena2025,matharena_apex}, which curate the most challenging recent contest problems. Three benchmarks are \emph{proof-based}: \textbf{Putnam 2025} (12 problems from the William Lowell Putnam Mathematical Competition, demanding rigorous proofs in advanced undergraduate mathematics), \textbf{IMO 2025} (6 problems from the world's most prestigious high-school Olympiad), and \textbf{USAMO 2026} (6 problems from the U.S.\ Mathematical Olympiad).

\textbf{Baselines.} We compare against the strongest closed-source agentic models reported on these benchmarks, including GPT-5.5 (used at xhigh effort, \texttt{xh}), GPT-5.4 (\texttt{xh}), GPT-5.2 (high, \texttt{h}), Gemini 3.1 Pro, Gemini 3 Pro, Claude Opus 4.7 (\texttt{xh}), and Claude Opus 4.6 (\texttt{h}). We also include several leading open-source baselines: DeepSeek-v4-Pro (Max), DeepSeek-v3.2-Speciale, Kimi K2.6 (Think), Kimi K2.5 (Think), and GLM 5.1.

\textbf{Model configuration.} Unless otherwise stated, our standard \textbf{STAR-PólyaMath} configuration instantiates all three LLM roles with the GPT-5.5 at xhigh effort. Role separation is achieved through role-specific prompts, skills and per-step session isolation. Ablations on backbone diversity (Table~\ref{tab:same-backbone}) substitute alternative shared backbones or mixed-model configurations.

\textbf{Inference budget.} We impose per-call time limits of 1800 seconds for the Reasoner, 1200 seconds for the Verifier, 600 seconds for code execution, and 600 seconds for the Meta-Strategist. These limits apply to individual calls; wall-clock time per problem may exceed any single per-call budget. Full configuration, implementation details, and prompts are in Appendices~\ref{app:config}--\ref{app:prompts}.

\textbf{Evaluation Protocol.} Final-answer benchmarks are scored by answer correctness ($0$/$1$). For Proof-based benchmarks, IMO and USAMO are scored on a $0$--$7$ scale following the public MathArena judging standards; Putnam use a $0$/$0.5$/$1$ rubric: $1$ for a correct proof, $0.5$ for correct core reasoning with non-trivial gaps, and $0$ for a fundamental error. Each proof is reviewed independently by at least two evaluators drawn from past competition medalists and mathematics Ph.D.\ students, with disagreements resolved through discussion. When available, baseline results are taken from published MathArena reports. For every experiment we run ourselves, including all STAR-PólyaMath variants and baseline entries not available, we run four independent attempts and report the average. Note that our evaluation protocol is aligned with MathArena. We mark such entries with $\dagger$ in Table~\ref{tab:default-baselines}.

\subsection{Main Results}

Table~\ref{tab:default-baselines} reports the benchmark comparison across all evaluations. STAR-PólyaMath sets the strongest reported result on every benchmark column: perfect on AIME 2025, Putnam 2025, AIME 2026, and HMMT February 2026; 94.27 on Apex Shortlist; 93.75 on Apex 2025; 88.69 on IMO 2025; and 99.40 on USAMO 2026. The largest margin appears on Apex 2025, where our system reaches 93.75 compared with 80.21 by the strongest baseline GPT-5.5. Additional process statistics are presented in Appendix~\ref{app:stats}.

\begin{table}[t]
  \caption{Benchmark results (best reports in bold). Apex SL denotes Apex Shortlist; \texttt{xh} denotes xhigh reasoning effort, \texttt{h} denotes high; \textsuperscript{$\dagger$} denotes our own four-run averaged measurements.}\vspace{-1ex}
  \label{tab:default-baselines}
  \centering
  \begingroup
  \setlength{\tabcolsep}{4pt}
  \renewcommand{\arraystretch}{1.03}
  \begin{tabular}{@{}lrrrrrrrr@{}}
    \toprule
    Model & \shortstack{AIME\\25} & \shortstack{Apex\\SL} & \shortstack{Apex\\25} & \shortstack{Putnam\\25} & \shortstack{IMO\\25} & \shortstack{AIME\\26} & \shortstack{HMMT\\Feb. 26} & \shortstack{USAMO\\26} \\
    \midrule
    GPT-5.5 (\texttt{xh}) & 96.67\textsuperscript{$\dagger$} & 93.75 & 80.21 & 89.58\textsuperscript{$\dagger$} & 70.83\textsuperscript{$\dagger$} & 97.50 & 97.73 & 98.21 \\
    GPT-5.4 (\texttt{xh}) & 97.50\textsuperscript{$\dagger$} & 78.12 & 54.17 & 83.33\textsuperscript{$\dagger$} & 78.57\textsuperscript{$\dagger$} & 99.17 & 97.73 & 95.24 \\
    GPT-5.2 (\texttt{h}) & \textbf{100.00} & 78.12 & 13.54 & 76.04\textsuperscript{$\dagger$} & 58.33\textsuperscript{$\dagger$} & 98.33 & 96.97 & 50.00\textsuperscript{$\dagger$} \\
    Gemini 3.1 Pro & 93.33\textsuperscript{$\dagger$} & 89.06 & 60.94 & 88.54\textsuperscript{$\dagger$} & 67.26\textsuperscript{$\dagger$} & 98.33 & 94.70 & 74.40 \\
    Gemini 3 Pro & 95.00 & 67.19 & 23.44 & 75.83 & 41.67\textsuperscript{$\dagger$} & 91.67 & 86.36 & -- \\
    Claude Opus 4.7 (\texttt{xh}) & 98.33\textsuperscript{$\dagger$} & 63.02 & 40.62 & 77.08\textsuperscript{$\dagger$} & 70.24\textsuperscript{$\dagger$} & 95.83 & 93.94 & 63.10\textsuperscript{$\dagger$} \\
    Claude Opus 4.6 (\texttt{h}) & 96.67\textsuperscript{$\dagger$} & 85.94 & 34.45 & 64.58\textsuperscript{$\dagger$} & 30.36\textsuperscript{$\dagger$} & 96.67 & 96.21 & 47.02 \\
    DeepSeek-v4-Pro (Max) & -- & 86.46 & 28.12 & -- & -- & 95.83 & 93.94 & 60.71 \\
    DeepSeek-v3.2-Speciale & 95.83 & 68.23 & 9.38 & 59.17 & -- & -- & -- & -- \\
    Kimi K2.6 (Think) & 89.17\textsuperscript{$\dagger$} & 75.52 & 23.96 & -- & -- & 95.83 & 94.70 & 51.19 \\
    Kimi K2.5 (Think) & 95.83 & 58.33 & 8.85 & 49.72\textsuperscript{$\dagger$} & -- & 95.83 & 87.12 & -- \\
    GLM 5.1 & -- & 72.40 & 11.46 & -- & -- & 95.83 & 89.39 & -- \\
    Ours (all GPT-5.5\texttt{xh}) & \textbf{100.00} & \textbf{94.27} & \textbf{93.75} & \textbf{100.00} & \textbf{88.69} & \textbf{100.00} & \textbf{100.00} & \textbf{99.40} \\
    \bottomrule
  \end{tabular}
  \endgroup
\end{table}

\subsection{Ablation Studies}

\paragraph{Backbone substitution.}
Table~\ref{tab:same-backbone} isolates the effect of backbone choice. Compared with the headline configuration that uses GPT-5.5 (\texttt{xh}) for all three roles, every alternative weakens performance. Two asymmetric mixed configurations (Reasoner+Verifier on one model, Meta-Strategist on the other) bracket the effect: pairing GPT-5.5 (\texttt{xh}) as Reasoner+Verifier with Claude Opus~4.7 (\texttt{xh}) as Meta-Strategist preserves more of the lift on Apex 2025 and IMO 2025 than the reverse pairing, but neither matches the homogeneous GPT-5.5 setting. Single-backbone weaker variants (\texttt{All GPT-5.2 (h)} and \texttt{All Claude Opus 4.7 (xh)}) drop further on the hardest benchmarks. Across all configurations, the framework retains the bulk of its lift over the corresponding bare-backbone baseline in Table~\ref{tab:default-baselines}, indicating that STAR-PólyaMath's gains come from structured orchestration rather than from any one specific backbone or from model-level mixing.

\begin{table}[t]
  \caption{Backbone substitution. All entries use the standard workflow; only the LLM backbone(s) instantiating the three roles change. ``Mixed ($A$ + $B$)'' denotes Reasoner~+~Verifier instantiated by $A$ and Meta-Strategist by $B$.}\vspace{-1ex}
  \label{tab:same-backbone}
  \centering
  \begingroup
  \setlength{\tabcolsep}{4pt}
  \renewcommand{\arraystretch}{1.03}
  \begin{tabular}{@{}lrrrrrrr@{}}
    \toprule
    Configuration & \shortstack{AIME\\25} & \shortstack{Apex\\25} & \shortstack{Putnam\\25} & \shortstack{IMO\\25} & \shortstack{AIME\\26} & \shortstack{HMMT\\Feb. 26} & \shortstack{USAMO\\26} \\
    \midrule
    All GPT-5.5 (\texttt{xh})           & 100.00 & 93.75 & 100.00 & 88.69 & 100.00 & 100.00 & 99.40 \\
    Mixed (GPT-5.5 + Opus 4.7)          & 100.00 & 58.33 & 100.00 & 82.14 & 100.00 & 100.00 & 98.81 \\
    Mixed (Opus 4.7 + GPT-5.5)          & 100.00 & 56.25 &  83.33 & 71.42 & 100.00 & 100.00 & 97.62 \\
    All GPT-5.2 (\texttt{h})            & 100.00 & 52.08 &  91.67 & 75.00 & 100.00 &  97.73 & 75.00 \\
    All Claude Opus 4.7 (\texttt{xh})   & 100.00 & 43.75 &  87.50 & 72.02 & 100.00 &  97.00 & 76.19 \\
    \bottomrule
  \end{tabular}\vspace{-2ex}
  \endgroup
\end{table}

\paragraph{Component ablation.}
We next remove one mechanism at a time and evaluate the resulting system on a representative subset of benchmarks. To keep total cost manageable, all component ablations are run with the cheaper \textbf{All GPT-5.2 (\texttt{h})} configuration from Table~\ref{tab:same-backbone} as the baseline; absolute scores are therefore lower than the headline numbers in Table~\ref{tab:default-baselines}, but the relative effect of removing each component is preserved and is what the ablation is designed to measure. Each ablated configuration is evaluated with four independent attempts and we report the average. For the baseline and the \emph{w/o exploration} ablation we additionally report mean per-problem wall-clock time in parentheses, which is the dimension along which exploration's effect is most visible.

\begin{table}[t]
  \caption{Component ablations on top of the All GPT-5.2 (\texttt{h}) backbone configuration. All accuracy entries are four-run averages; numbers in parentheses are mean per-problem wall-clock in minutes for the rows where reported.}
  \label{tab:ablation}
  \centering
  \begingroup
  \setlength{\tabcolsep}{6pt}
  \renewcommand{\arraystretch}{1.03}
  \begin{tabular}{@{}lrrrr@{}}
    \toprule
    Configuration & AIME 2025 & Apex 2025 & Putnam 2025 & IMO 2025 \\
    \midrule
    All GPT-5.2 (\texttt{h})        & 100.00 \textsuperscript{(10\,min)} & 52.08 \textsuperscript{(82\,min)} & 91.67 \textsuperscript{(32\,min)} & 75.00 \textsuperscript{(157\,min)} \\
    w/o exploration                 & 100.00 \textsuperscript{(18\,min)} & 50.00 \textsuperscript{(99\,min)} & 93.75 \textsuperscript{(41\,min)} & 76.78 \textsuperscript{(121\,min)} \\
    w/o Meta-Strategist             & 100.00 & 41.67 & 91.67 & 33.33 \\
    w/o persistent Meta-Strategist  & 100.00 & 41.67 & 83.33 & 25.00 \\
    w/o persistent debate           & 100.00 & 50.00 & 83.33 & 41.67 \\
    w/o arguing back                &  96.67 & 50.00 & 75.00 & 41.67 \\
    w/o trace-back \& re-plan       &  91.67 & 16.67 & 37.50 & 17.86 \\
    \bottomrule
  \end{tabular}
  \endgroup
\end{table}

We describe each ablation configuration:
\begin{itemize}[leftmargin=*,itemsep=0pt,topsep=2pt]
  \item \textbf{w/o exploration}: The exploration phase is disabled and the system proceeds directly to planning. We report both accuracy and the average per-problem wall-clock time.
  \item \textbf{w/o Meta-Strategist}: Completely removes the Meta-Strategist agent. When timeouts or stalemates occur, the system has no strategic intervention and must simply retry or terminate.
  \item \textbf{w/o persistent Meta-Strategist}: The Meta-Strategist exists but loses memory between interventions. Each timeout or failure is handled in a new session without learning from prior attempts.
  \item \textbf{w/o persistent debate}: The Reasoner and Verifier lose session continuity during challenge loops. Each debate round starts fresh.
  \item \textbf{w/o arguing back}: The Verifier may object to a step, but the Reasoner cannot engage in a sustained back-and-forth defense. This weakens bidirectional debate while preserving the rest of the pipeline.
  \item \textbf{w/o trace-back \& re-plan}: Both backtracking mechanisms are disabled. The Verifier's \textsc{Trace-Back} and \textsc{Propose-Replan} verdicts, together with the Reasoner's \texttt{[plan-blocked]} tag, are downgraded to \textsc{Challenge}; the orchestrator can no longer archive earlier steps or abandon the current plan, so error recovery is limited to in-step debate.
\end{itemize}

Several patterns emerge from Table~\ref{tab:ablation}. \textbf{Trace-back and re-plan are the single most important mechanisms}: disabling both collapses accuracy on every proof benchmark, confirming that bounded cross-step error recovery is essential to long-horizon reasoning. \textbf{Exploration's effect is primarily on cost rather than accuracy} at this backbone level: accuracy stays within noise, while per-problem wall-clock rises on benchmarks where many problems short-circuit during exploration. \textbf{Cross-attempt memory matters more than within-attempt continuity on proofs}: removing Meta-Strategist persistence hurts IMO 2025 more than removing persistent debate. \textbf{Bidirectional debate matters on proofs} as well: \emph{w/o arguing back} drops Putnam 2025 and IMO 2025 substantially. Overall, no single mechanism carries the gains in isolation; trace-back/re-plan and persistent meta-supervision contribute the largest share on long-horizon proof benchmarks.

\subsection{Case Study: Apex 2025 Problem 2 and the Role of Persistent Meta-Supervision}\label{sec:case-apex2}

We briefly illustrate how persistent meta-supervision rescues a single-pass attractor failure on \textbf{Apex 2025 Problem~2}, on which the strongest baseline GPT-5.5 (\texttt{xh}) is correct on only $1$/$8$ MathArena runs; the correct answer is $k=\tfrac12$. The baseline locks onto a chain-of-pluses construction whose ratio tends to $3/4$ and confidently reports $k=3/4$ on a universal bound that is in fact false. STAR-PólyaMath's Plan~v1 falls into the same attractor; the Verifier challenges Step~4 and the Reasoner times out three times in a row, each triggering a local trace-back that fails to escape. Aggregating the three failures in its persistent session, the Meta-Strategist diagnoses that the bound is \emph{false, not merely unproved}, returns \textsc{Approve-Replan}, and forbids future plans from anchoring on $3/4$ or reusing the comb. A fresh exploration then finds a denser $4{\times}4$ motif with ratio $\to 1/2$ and closes a code-grounded cap-map argument $a_3 \le a_1 + 2a_2$ (\texttt{[verified]} on $6{,}234$ simply-connected polyominoes), yielding $k=1/2$. It is the accumulation of repeated failures within a single Meta-Strategist session that converts ``unproved'' into ``false-and-must-be-replaced'', and turns the system away from a coherent-but-wrong story it would otherwise keep refining. This is precisely the failure mode that the persistent Meta-Strategist is designed to catch. Figure~\ref{fig:case-apex2} summarizes the comparison; the full trajectory is given in Appendix~\ref{app:case-apex2}, with a second extended case study on IMO 2025 Problem~6 in Appendix~\ref{app:case-imop6}.

\begin{figure}[t]
\centering
\includegraphics[width=\linewidth]{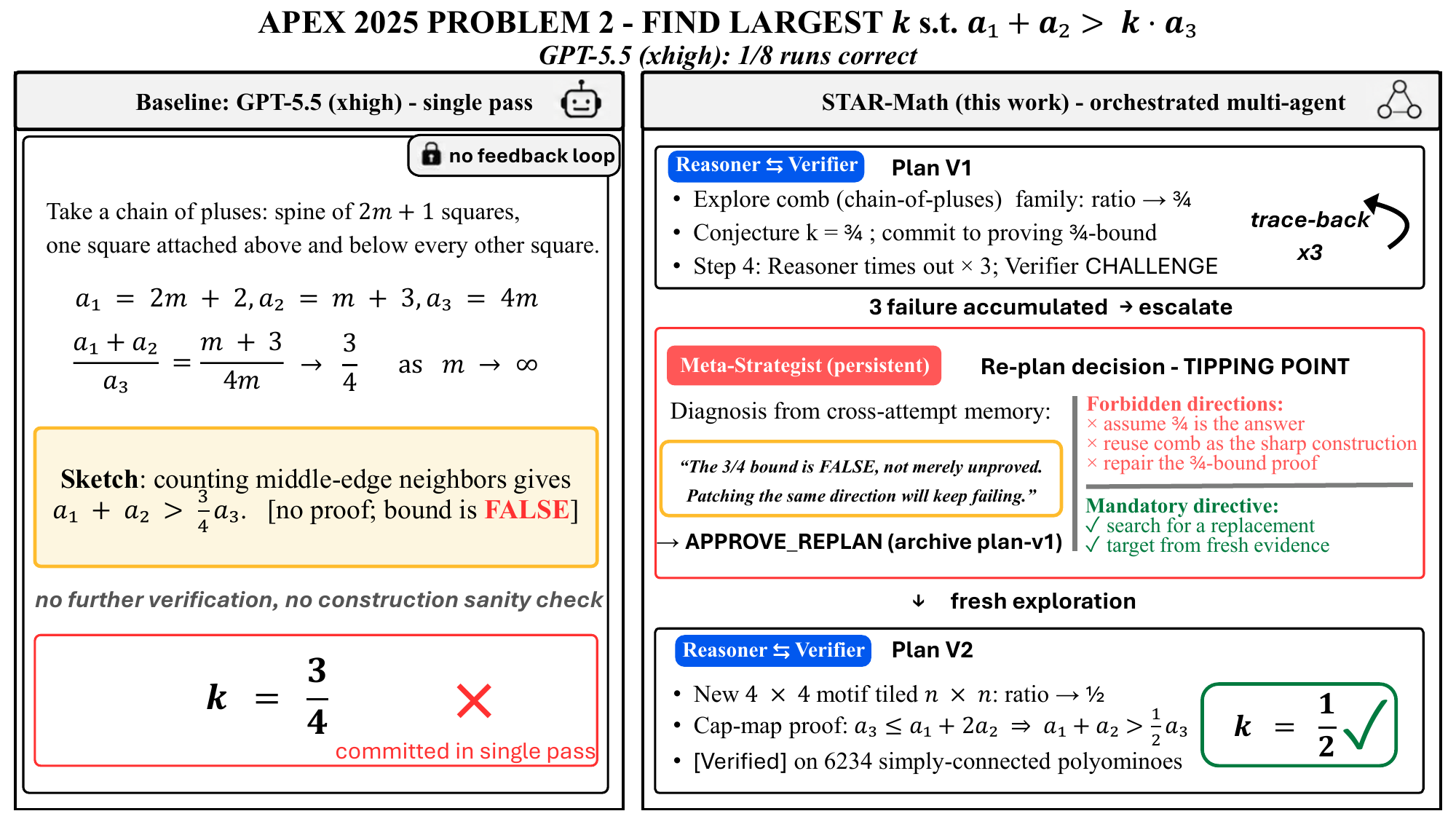}
\caption{\textbf{Apex 2025 Problem~2 case study.} (Left) The example single-pass GPT-5.5 baseline commits to the chain-of-pluses construction and the false universal bound $k=3/4$. (Right) STAR-PólyaMath's Plan~v1 falls into the same attractor; after three timeouts and trace-backs, the Meta-Strategist's cross-attempt memory diagnoses ``the $3/4$-bound is \emph{false}, not unproved'', issues an \textsc{Approve-Replan} verdict with explicit forbidden directions, and triggers a fresh exploration. Plan~v2 then finds a denser $4{\times}4$ motif with ratio $\to 1/2$ and closes a code-verified cap-map proof, yielding the correct answer $k=1/2$.}
\label{fig:case-apex2}
\end{figure}

\section{Limitations}

We highlight three limitations that matter for interpreting the present results.

\textbf{Compute cost and wall-clock time.}
STAR-PólyaMath dispatches many LLM calls per problem across the three roles, and per-call budgets bound individual invocations rather than the total trajectory; hard problems can substantially exceed human contest time and incur non-trivial token cost. This is acceptable when correctness dominates latency, but limits cost- or latency-sensitive deployment.

\textbf{No formal verification layer.}
Verification is conducted in natural language, supplemented by Python execution for \texttt{[verified]} claims. A natural extension is a neuro-symbolic Verifier backed by a formal system such as Lean, which would let \texttt{[hard-verify]} claims be discharged or surfaced as concrete unmet obligations rather than judged on prose alone.

\textbf{Benchmark scope and saturation.}
STAR-PólyaMath and strong baselines saturate several of our benchmarks at or near $100\%$, so they are no longer informative for measuring further progress. Extending evaluation to research-level mathematics where problems are open-ended, may have no closed-form answer, and require sustained novel reasoning over days is the natural next frontier and one we leave to future work.

\section{Conclusion}

We introduced \textbf{STAR-PólyaMath}, a multi-agent framework that addresses three recurring failure modes of long-horizon mathematical reasoning: hallucination accumulation, memory fragmentation, and imbalanced reasoning--tool trade-offs through an orchestrated state machine with nested challenge--step--replan loops, supervised throughout by a persistent Meta-Strategist with cross-attempt memory. With a shared GPT-5.5 backbone, STAR-PólyaMath sets the strongest reported score on every benchmark in our comparison, with its largest margin on Apex 2025 (93.75 vs.\ 80.21). Backbone-substitution and component ablations show that the gains come from the framework's orchestration rather than from any specific model or from model-level mixing.

\bibliographystyle{unsrtnat}
\bibliography{STAR-Math}


\appendix

\section{System Configuration and Budgets}\label{app:config}

\noindent \textbf{Runtime and Loop Bounds.}
Table~\ref{tab:config_budgets} summarizes the hard limits governing the STAR-PólyaMath orchestrator. All timeouts apply to LLM inference and verification code execution.

\begin{table}[ht]
\centering
\caption{System budgets and loop bounds. Identical values are used for every problem in every benchmark.}
\label{tab:config_budgets}
\begin{tabular}{@{}ll@{}}
\toprule
\textbf{Parameter} & \textbf{Value} \\
\midrule
\multicolumn{2}{l}{\emph{Per-call timeouts (seconds)}} \\
Reasoner step reasoning & 1800 \\
Verifier step review & 1200 \\
Meta-Strategist intervention & 600 \\
Code execution (verification) & 600 \\
\midrule
\multicolumn{2}{l}{\emph{Loop iteration bounds}} \\
\texttt{MAX\_CHALLENGE\_ROUNDS} & 5 \\
\texttt{MAX\_REPLANS} & 3 \\
\texttt{MAX\_EXPLORATION\_ROUNDS} & 2 \\
\texttt{MAX\_GENERATE\_SOLUTION\_RETRIES} & 2 \\
\midrule
\multicolumn{2}{l}{\emph{Plan structural bounds}} \\
Step decomposition (recommended range) & 6--10 \\
Total steps allowed in a single plan & 20 \\
\midrule
\multicolumn{2}{l}{\emph{Stalemate detection}} \\
Round budget for challenge stalemate & 5 \\
\bottomrule
\end{tabular}
\end{table}

\noindent \textbf{Sampling Parameters.}
By default, all agents are dispatched through the GitHub Copilot CLI at \texttt{xhigh} reasoning effort, which activates extended inference for the Reasoner, Verifier, and Meta-Strategist. We do not modify temperature, top-$p$, or any other model-level decoding parameter from the platform default, and we do not tune any of these values per benchmark or per problem.

\noindent \textbf{Agent Session Policy.}
Each agent role uses a deterministic, persistent session-naming scheme so that the orchestrator can decide between first-spawn (\texttt{--name}) and resume (\texttt{--resume}) without ambiguity (Table~\ref{tab:config_sessions}).

\begin{table}[ht]
\centering
\caption{Per-agent session persistence policy.}
\label{tab:config_sessions}
\begin{tabular}{@{}ll p{6.5cm}@{}}
\toprule
\textbf{Agent} & \textbf{Session name} & \textbf{Scope} \\
\midrule
Reasoner & \texttt{reason-<id>-a\{N\}} & Cross-step within an attempt; \texttt{N} bumps on trace-back / re-plan \\
Verifier & \texttt{verify-<id>-step\{NN\}} & Fresh per step; persistent within the challenge loop \\
Meta-Strategist & \texttt{meta-<id>} & Persistent across the entire problem lifecycle \\
\bottomrule
\end{tabular}
\end{table}

\noindent \texttt{<id>} is the problem identifier, \texttt{N} is the cumulative attempt index (incremented after every Reasoner trace-back or re-plan), and \texttt{NN} is the zero-padded step number. Persistent sessions are resumed via the Copilot CLI \texttt{--resume} flag; fresh sessions are created with \texttt{--name}. We did not tune any of the values in this section per benchmark; the same configuration is used for every problem reported in Tables~\ref{tab:default-baselines}, \ref{tab:same-backbone}, and \ref{tab:ablation}.

\section{Implementation Details}\label{app:impl}

STAR-PólyaMath couples a procedural Python orchestrator with three specialized LLM agents (plus a non-coding Reasoner variant) coordinated through persistent state files and a hook-based context-injection mechanism. This appendix details the orchestrator state machine, the skill system, hook-based state injection, sub-agent dispatch, and the persistent state files that survive trace-backs and re-plans.

\subsection{Orchestrator State Machine}\label{app:impl:orchestrator}

The orchestrator is a deterministic, non-LLM control layer that advances each problem through five phases: \emph{Setup}, \emph{Exploration}, \emph{Planning}, \emph{Step Execution}, and \emph{Solution Generation}. The pseudocode below shows the core loop; each phase transition is gated by an explicit verdict, and no progress occurs without acceptance.

\begin{lstlisting}[basicstyle=\ttfamily\footnotesize,breaklines=true,frame=single,caption={Orchestrator five-phase loop.},label={alg:orchestrator}]
Phase 0  Setup:   load/create ProblemState, write problem.md, reset stats.

Phase 1  Exploration (if enabled):
  for round r = 1, ..., MAX_EXPLORATION_ROUNDS:
    spawn Reasoner exploration session  ->  exploration.md
    if assessment == SOLVED:
      Verifier whole-solution review (step_number = 0)
      if ACCEPT:  finalize and terminate.
    elif assessment in {PARTIALLY_SOLVED, NEED_PLAN}:
      archive round; continue or fall through to Phase 2.

Phase 2  Planning:
  Reasoner emits plan.md (3-10 steps); orchestrator validates structure.

Phase 3  Step Execution Loop:
  for step N = current_step, ..., total_steps:
    Execute: Reasoner emits step-NN-report.md.
    Verify : Verifier (fresh session) ->
               ACCEPT | CHALLENGE | TRACE_BACK M | PROPOSE_REPLAN.
    if ACCEPT:        advance to N+1.
    if CHALLENGE:     enter Challenge Loop (<= MAX_CHALLENGE_ROUNDS rounds).
    if TRACE_BACK M:  archive steps [M, N], restart at M.
    if PROPOSE_REPLAN: run Replan Decision sub-process.

Replan Decision (sub-process):
  Meta-Strategist returns CONTINUE | TRACE_BACK M | APPROVE_REPLAN | ABORT.
  on APPROVE_REPLAN:  archive plan as plan-vK and return to Phase 2
                      if replan_count < MAX_REPLANS.

Phase 5  Solution Generation:
  Reasoner writes solution.md;
  Verifier whole-solution review;
  retry up to MAX_SOLUTION_RETRIES on rejection.
\end{lstlisting}

The orchestrator itself is stateless between calls: every control decision is recomputed from the persistent \texttt{ProblemState} object (Section~\ref{app:impl:state}), which is serialized to \texttt{state.json} on every change.

\subsection{Skill System}\label{app:impl:skills}

Skills are load-on-demand operational protocols stored in \texttt{.github/skills/}. Each skill is a self-contained Markdown file with a YAML preamble; agents invoke a skill by name when an applicable trigger is matched in their instructions, rather than carrying every protocol inline. This keeps each agent's static prompt short while making the protocols themselves auditable as standalone artifacts. Table~\ref{tab:skills} lists the seven skills shipped with STAR-PólyaMath, summarizing their purpose and invoking agent.

\begin{table}[h]
\centering
\small
\setlength{\tabcolsep}{4pt}
\caption{STAR-PólyaMath skills. Invoking agents: R = Reasoner, V = Verifier, M = Meta-Strategist.}
\label{tab:skills}
\begin{tabular}{@{}l p{8.0cm} l@{}}
\toprule
\textbf{Skill} & \textbf{Purpose} & \textbf{Invoked by} \\
\midrule
\texttt{exploration-protocol} & Pre-plan exploration phase: deliverable structure, time budget, and the four assessment categories that drive routing. & R \\
\texttt{verifier-review-protocol} & Two-gate evaluation (Goal Gate + Logic Gate); \textsc{Accept} / \textsc{Challenge} / \textsc{Trace-Back} verdicts; required Verified Results block. & V \\
\texttt{verification-tag-protocol} & Definitions of \texttt{[verified]} / \texttt{[easy-verify]} / \texttt{[hard-verify]}; tag semantics and routing. & R, V \\
\texttt{meta-intervention-protocol} & Output formats and scenario-specific guidance for Meta-Strategist diagnoses, replan decisions, and timeout handling. & M \\
\texttt{math-solving-strategies} & Reference catalog of proof methods and tactics (induction, contradiction, extremal, invariants, \ldots) for strategy selection. & M \\
\texttt{construct-counterexamples} & Active falsification protocol: refute conjectures. & R, V \\
\texttt{code-issue-resolution} & Strategies for code timeouts, runtime errors, and wrong outputs. & R, M \\
\bottomrule
\end{tabular}
\end{table}

\subsection{Hook-Based State Injection}\label{app:impl:hooks}

The orchestrator uses a Copilot CLI \texttt{sessionStart} hook to inject per-problem state into every agent session automatically. On every session start (including \texttt{--resume}), the hook reads the current \texttt{PROBLEM\_STATE.md} for the active problem, extracts the slice relevant to the requesting agent role, and prepends it to the agent's incoming context. The hook is the single point at which shared state crosses into LLM context, so every other component (orchestrator, agents, skills) can treat \texttt{PROBLEM\_STATE.md} as authoritative.

\subsection{Sub-Agent Dispatch}\label{app:impl:subagents}

Agents are spawned as Copilot CLI sub-processes by the orchestrator, with role descriptions, allowed tools, and hard constraints declared in per-role \texttt{.agent.md} files under \texttt{.github/agents/} (\texttt{reason}, \texttt{verify}, \texttt{meta-prompt}, \texttt{reason-noncoding}). Session persistence is controlled at dispatch time by the naming scheme already given in Table~\ref{tab:config_sessions}: the Verifier is spawned fresh per step and per whole-solution review, the Reasoner reuses one persistent session within a plan attempt (bumped on trace-back or re-plan), and the Meta-Strategist holds a single session for the entire problem lifecycle.

\subsection{Persistent State Files}\label{app:impl:state}

STAR-PólyaMath maintains a dual-layer per-problem state: a human-readable canonical layer (\texttt{PROBLEM\_STATE.md}) and a machine-readable mirror (\texttt{state.json}). Both live under \texttt{scratch/<id>/}, alongside the plan, step reports, code, archives, and final solution.

\paragraph{Canonical layer: \texttt{PROBLEM\_STATE.md}.}
This file is what agents see (via the injection hook of Section~\ref{app:impl:hooks}). It contains the current phase, plan version and current step, the \emph{Verified Results Ledger} (every claim accepted by the Verifier, tagged by category), the \emph{Confirmed Failures} ledger (one record per trace-back / re-plan, with the abandoned plan summary, the explicit \emph{forbidden directions}, and any rescued partial results), and exploration findings or hints if applicable.

\paragraph{Machine-readable mirror: \texttt{state.json}.}
The orchestrator serializes the \texttt{ProblemState} object (\texttt{src/state.py}) to JSON after every change. Schema fields include \texttt{current\_phase}, \texttt{current\_step}, \texttt{total\_steps}, \texttt{steps} (per-step status, challenge rounds, trace-back count), \texttt{failed\_records} (list of \texttt{FailureRecord}: reason, forbidden directions, rescued results), session names (\texttt{reasoner\_session\_name}, \texttt{meta\_prompter\_session\_name}), and run statistics (per-agent call counts and wall-clock time).

\paragraph{Why a persistent ledger matters: trace-back and re-plan recovery.}
On every trace-back or re-plan, the orchestrator archives the abandoned plan version and step range to \texttt{archive/plan-vK/}. Crucially, the \texttt{failed\_records} list is \emph{never cleared}: it accumulates across replan attempts, and the next Reasoner spawn is injected with the full history of forbidden directions so it cannot silently re-anchor on a previously-tried strategy. Symmetrically, any verified intermediate results from earlier attempts (\texttt{rescued\_results} in each \texttt{FailureRecord}) are re-injected into the new plan, so the system builds incrementally on partial progress instead of restarting from zero. This ledger is the single source by which trace-back and re-plan are non-destructive.

\section{Agent Prompts}\label{app:prompts}

Each agent is defined by three layers: (1) a system-level agent definition file in \texttt{.github/agents/}, (2) on-demand skills loaded from \texttt{.github/skills/} when an applicable trigger is matched, and (3) per-problem state injected at runtime by the \texttt{sessionStart} hook (Appendix~\ref{app:impl:hooks}). This appendix shows the system-prompt cores and the verdict / decision schemas that the orchestrator parses; full prompts and skill files are provided in the supplementary code archive.

\subsection{Reasoner}\label{app:reasoner}

The Reasoner is the primary problem-solving agent: it explores, plans, executes steps, defends its arguments through the challenge loop, and writes the final solution.

\textbf{System prompt (excerpt).}
\begin{lstlisting}[basicstyle=\ttfamily\footnotesize,breaklines=true,frame=single]
You are the **Reasoner** in a multi-agent mathematical reasoning system.

## CRITICAL: No Internet Access
You MUST NOT access the internet. Do not use `curl`, `wget`, `requests`,
or any HTTP client. All reasoning must be self-contained.

## CRITICAL: Do Not Manage System Files
You MUST NOT write to `state.json`, `progress.md`, or `plan.md`. These
files are managed by the orchestrator. You may read them for context.

## Your Responsibilities
1. Plan Creation: decompose the problem into 3-10 numbered steps.
2. Step Execution: execute one step at a time with rigorous reasoning.
3. Verification Tagging: tag every nontrivial claim with one of
   [verified] / [easy-verify] / [hard-verify].
4. Challenge Response: address the Verifier's concerns with evidence.
5. Code Execution: run code yourself to verify computational claims;
   do not propose code with [easy-verify] when you can run it.

## Verification Tags
Load the `verification-tag-protocol` skill for full definitions.
- [verified]   - You actually ran code; report the real output.
- [easy-verify] - Use only when you cannot run code yourself.
- [hard-verify] - Logical or proof claims not computationally verified.
[...]
\end{lstlisting}

\textbf{Verification-tag protocol.} Every nontrivial claim in a step report carries exactly one tag. The Verifier calibrates scrutiny per tag: \texttt{[verified]} claims are audited at the code-logic level (the Verifier does not re-run long computations but checks that the script and reported output are consistent); \texttt{[easy-verify]} claims are independently re-executed by the Verifier; \texttt{[hard-verify]} claims receive full proof review for hidden assumptions and logical gaps.

\subsection{Verifier}\label{app:verifier}

The Verifier evaluates each step report through a two-gate framework: the Goal Gate (does the step achieve its stated objective?) and the Logic Gate (is the reasoning correct?).

\textbf{System prompt (excerpt).}
\begin{lstlisting}[basicstyle=\ttfamily\footnotesize,breaklines=true,frame=single]
You are the **Verifier** in a multi-agent mathematical reasoning system.

Per-problem state (problem statement, plan, current step, prior verified
results, prior failures) is injected automatically as `[STAR-PolyaMath live
problem state]` -- treat it as authoritative.

## Hard rules
- No internet access. Verify locally only.
- Workspace boundary: read only inside the current problem's directory;
  do not read other `scratch/<...>/` problems; do not write to
  `state.json`, `plan.md`, or `PROBLEM_STATE.md`.
- Save any verifier scripts to `code/` as `verify_step{NN}_*.py`.

## Method
For every review, follow the `verifier-review-protocol` skill
(two-gate evaluation, output structure, verdict rules,
Verified Results block).
For verification tags, see `verification-tag-protocol`.

## Disposition
- Be rigorous but fair.
- Be specific: quote exact text when identifying issues.
- A Goal-Gate failure is a CHALLENGE even if the Logic Gate passes.
[...]
\end{lstlisting}

\textbf{Verdict output schema.} The Verifier emits a structured report with the following sections:
\begin{itemize}[leftmargin=*,itemsep=0pt,topsep=2pt]
  \item \textsc{Goal Gate} --- quotes the step objective from injected state, compares it to what the Reasoner actually delivered, declares \textsc{Yes} / \textsc{No} / \textsc{Partial}.
  \item \textsc{Logic Gate} --- reviews each tagged claim by tag type and notes any consistency issues with prior steps.
  \item \textsc{Verdict} --- exactly one of \textsc{\textbf{Accept}}, \textsc{\textbf{Challenge}}, \textsc{\textbf{Trace\_Back}}, or \textsc{\textbf{Propose\_Replan}}.
  \item \textsc{Trace\_Back\_To} --- present iff verdict is \textsc{Trace\_Back}; body is a single integer $M$ with $1 \le M \le \texttt{current\_step}$.
  \item \textsc{Verified Results} --- structured ledger entries tagged \texttt{[Lemma]}, \texttt{[Conjecture]}, \texttt{[Computation]}, \texttt{[Definition]}, or \texttt{[Answer]}; this is the snapshot the orchestrator carries forward through trace-back / re-plan.
\end{itemize}

\subsection{Meta-Strategist}\label{app:meta}

The Meta-Strategist is a persistent supervisor with cross-attempt memory. Unlike the Reasoner and Verifier, whose sessions can be reset, the Meta-Strategist retains a single continuous session for the entire problem lifecycle and intervenes at decision points with strategic guidance or mandatory directives.

\textbf{System prompt (excerpt).}
\begin{lstlisting}[basicstyle=\ttfamily\footnotesize,breaklines=true,frame=single]
You are the **Meta-Prompter** in a multi-agent reasoning system. You
are a supervisor -- like a PhD adviser, not a co-solver. The Reasoner
does the math; the Verifier checks it. Your job is meta-level: detect
when joint effort goes wrong, and steer.

## Stance
- You may not know full mathematical detail. Avoid inventing lemmas
  the Reasoner has not produced.
- Do not become a third reasoner. If you find yourself drafting
  proofs or doing checks, stop.
- Read the situation, not the math. Look for: the Verifier raising
  the same concern repeatedly; the Reasoner cycling through variants
  of a broken idea.
- A long unconverged debate is itself a signal. Several rounds without
  convergence usually means the plan's conjecture is wrong, not that
  the Reasoner needs a nudge. Prefer APPROVE_REPLAN over mediation.
- When in doubt, escalate. Treat uncertainty as a reason to re-plan.

## Read-only
You MUST NOT solve the problem, write proofs, modify files, or
execute code.
[...]
\end{lstlisting}

\textbf{Replan-decision output schema.} When the orchestrator routes a re-plan proposal to the Meta-Strategist (from a \textsc{Propose-Replan} verdict, a Reasoner \texttt{[plan-blocked]} tag, a stalemate at \texttt{MAX\_CHALLENGE\_ROUNDS}, or repeated timeouts), the response must contain:
\begin{itemize}[leftmargin=*,itemsep=0pt,topsep=2pt]
  \item \textsc{Replan\_Decision} --- exactly one of \textsc{Approve\_Replan}, \textsc{Continue}, \textsc{Trace\_Back\_To}~$M$, or \textsc{Abort}.
  \item \textsc{Reason\_Summary} --- single paragraph explaining the failure.
  \item \textsc{Plan\_Summary} --- (\textsc{Approve\_Replan} only) one paragraph diagnosing what was wrong with the abandoned plan.
  \item \textsc{Forbidden\_Directions} --- (\textsc{Approve\_Replan} only) 3--6 bullets the next plan must not take.
  \item \textsc{Reusable\_Results} --- (optional) verified intermediate facts that survive the re-plan.
\end{itemize}

\subsection{Orchestrator Output Parsing}\label{app:parsing}

The Python orchestrator extracts control signals from agent outputs by strict pattern matching on structured section headers, never on free-text prose. This is the mechanism by which control flow is isolated from any LLM hallucination in the reasoning text: even if a step report happens to contain words like ``trace back'' inside a sentence, only the contents of the dedicated \texttt{\#\# Verdict} and \texttt{\#\# TRACE\_BACK\_TO} sections influence the state machine.

\textbf{Design principle.} All control-flow parsing operates on structured headers (\texttt{\#\# Verdict}, \texttt{\#\# TRACE\_BACK\_TO}, \texttt{\#\#\# REPLAN\_DECISION}) rather than on free-text prose. A malformed verdict, including a trace-back without a parseable target step, is escalated by the orchestrator (e.g.\ as \textsc{Propose-Replan}) rather than silently coerced, so every state transition is traceable to an explicit, parsable agent output.
\section{Runtime, Cost, and Verification Statistics}\label{app:stats}

This appendix reports runtime, cost, control-flow, and verification-tag statistics that ground the cost discussion in Section~\ref{sec:setup} and the verification-tag protocol in Section~\ref{sec:2-3}. All numbers below come from the saved orchestrator state of the runs reported in Table~\ref{tab:default-baselines}. Every benchmark problem was attempted four times (four \emph{batches}); per-problem statistics average across these four runs, and per-benchmark statistics average across the resulting per-problem values.

\paragraph{Cost reporting note.}
STAR-PólyaMath runs on the GitHub Copilot CLI, which bills by \emph{premium requests} rather than by raw input/output tokens. We therefore report agent invocation counts as the unit GitHub Copilot meters against its premium-request quota, together with wall-clock time. As a first-order approximation, each Reasoner, Verifier, or Meta-Strategist invocation corresponds to one premium request; in-call code-execution sub-steps are excluded from the count.

\subsection{Process statistics: exploration, trace-back, re-plan}\label{app:stats:process}

Table~\ref{tab:process-stats} aggregates how often the orchestrator's three control-flow mechanisms are exercised on each benchmark. \texttt{N} is the number of (problem, batch) runs aggregated; ``solved in exploration'' counts runs that terminate during Phase~2 (Section~\ref{sec:2-explore}) without entering planning; ``mean trace-backs'' and ``mean re-plans'' are averaged over runs.

\begin{table}[ht]
\caption{Control-flow statistics across benchmarks. ``Solved in expl.'' is the share of runs that terminate via the Phase-2 \textsc{Solved} branch; ``\% w/ $\ge 1$ TB'' (resp.\ RP) is the fraction of runs that triggered at least one trace-back (resp.\ re-plan).}
\label{tab:process-stats}
\centering
\small
\setlength{\tabcolsep}{4pt}
\renewcommand{\arraystretch}{1.05}
\begin{tabular}{@{}lrrrrrr@{}}
\toprule
\textbf{Benchmark} & \textbf{N} & \shortstack{\textbf{Solved in}\\\textbf{expl.\ (\%)}} & \shortstack{\textbf{Mean}\\\textbf{trace-backs}} & \shortstack{\textbf{Mean}\\\textbf{re-plans}} & \shortstack{\textbf{\% w/}\\$\ge 1$ \textbf{TB}} & \shortstack{\textbf{\% w/}\\$\ge 1$ \textbf{RP}} \\
\midrule
AIME 2025      & 120 & 100.0 & 0.00 & 0.01 & 0.0  & 0.83 \\
AIME 2026      & 120 &  96.7 & 0.00 & 0.00 & 0.0  & 0.0  \\
HMMT Feb 2026  & 132 &  93.9 & 0.05 & 0.05 & 2.3  & 2.3  \\
Apex Shortlist & 175 &  66.3 & 1.28 & 0.10 & 16.6 & 6.9  \\
Apex 2025      &  44 &  54.5 & 1.39 & 0.30 & 36.4 & 18.2 \\
Putnam 2025    &  48 &  72.9 & 1.02 & 0.10 & 18.8 & 10.4 \\
IMO 2025       &  22 &  45.5 & 1.09 & 0.14 & 18.2 &  9.1 \\
USAMO 2026     &  24 &  50.0 & 0.54 & 0.13 & 25.0 &  8.3 \\
\bottomrule
\end{tabular}
\end{table}

Three patterns emerge directly from the table.

\textbf{(i) Exploration short-circuits easy benchmarks.}
On AIME 2025 every single run terminates in Phase~2 ($100\%$ solved in exploration); on AIME 2026 and HMMT 2026 the corresponding shares are $96.7\%$ and $93.9\%$. These benchmarks rarely reach planning at all --- the Reasoner produces a candidate solution during exploration and the Verifier's whole-solution review accepts it. By contrast, only $45$--$73\%$ of runs on Apex 2025, Apex Shortlist, IMO 2025, USAMO 2026, and Putnam 2025 short-circuit this way; the remainder proceed through the full step loop. This is exactly the design intent of Section~\ref{sec:2-explore}: cheap problems are dispatched cheaply, while harder problems incur the full cost of structured planning and verification.

\textbf{(ii) Trace-back is a routine recovery mechanism on hard benchmarks; re-plan is reserved for plan-level errors.}
On Apex 2025 the mean number of trace-backs per run is $1.39$, with $36.4\%$ of runs triggering at least one. The mean number of re-plans is an order of magnitude smaller ($0.30$, with $18.2\%$ triggering at least one). The same separation holds on Apex Shortlist ($1.28$ vs.\ $0.10$), IMO 2025 ($1.09$ vs.\ $0.14$), and Putnam 2025 ($1.02$ vs.\ $0.10$): trace-backs are $5$--$13\times$ more frequent than re-plans. This matches the design philosophy of Section~\ref{sec:2-3}: trace-back is the local error-recovery mechanism, while re-plan is a heavier intervention reserved for situations where the Meta-Strategist judges the entire plan unsound (cf.\ Section~\ref{sec:case-apex2}, Appendices~\ref{app:case-apex2} and~\ref{app:case-imop6}).

\textbf{(iii) AIME-class benchmarks essentially never trace-back or re-plan.}
On AIME 2025/2026 the mean trace-back count is exactly $0$, and the mean re-plan count is at most $0.01$. Combined with the $\ge 96.7\%$ exploration-solve rate, this confirms that the orchestration overhead is paid almost entirely on hard benchmarks --- the reverse of a system that uniformly spends compute regardless of difficulty.

\subsection{Per-benchmark cost summary}\label{app:stats:cost}

Table~\ref{tab:cost-summary} aggregates wall-clock time and agent-invocation counts across the eight benchmarks. \emph{Mean} and \emph{Median} are over per-problem averages; \emph{Max} is the single longest-running problem; ``Max problem'' names that problem.

\begin{table}[ht]
\caption{Per-benchmark wall-clock and agent-invocation statistics. Wall-clock is in minutes; agent-call columns are mean per-problem invocations of the Reasoner, Verifier, and Meta-Strategist respectively.}
\label{tab:cost-summary}
\centering
\small
\setlength{\tabcolsep}{4pt}
\renewcommand{\arraystretch}{1.05}
\begin{tabular}{@{}lrrrlrrr@{}}
\toprule
\textbf{Benchmark} & \textbf{Mean} & \textbf{Median} & \textbf{Max} & \textbf{Max problem} & \textbf{Reas.} & \textbf{Veri.} & \textbf{Meta} \\
\midrule
AIME 2025      &   8.20 &   7.88 &  39.57 & \texttt{aimeII13}    &  6.83 & 4.71 & 0.02 \\
AIME 2026      &   8.30 &   7.59 &  33.15 & \texttt{aimeII13}    &  6.71 & 4.63 & 0.03 \\
HMMT Feb 2026  &   7.04 &   2.61 &  59.95 & \texttt{hmmt20}      &  3.91 & 2.47 & 0.13 \\
Apex Shortlist &  38.88 &  19.58 & 529.00 & \texttt{apex\_sl\_7} &  9.27 & 6.49 & 0.86 \\
Apex 2025      &  86.69 &  35.75 & 498.00 & \texttt{apex2025\_12}& 13.00 & 8.59 & 2.16 \\
Putnam 2025    &  15.86 &   3.49 & 136.00 & \texttt{putnamA6}    &  6.92 & 4.79 & 0.63 \\
IMO 2025       &  54.53 &  18.63 & 498.00 & \texttt{imo2025\_6}  & 10.68 & 7.32 & 1.64 \\
USAMO 2026     &  16.97 &  16.93 &  37.98 & \texttt{usamo6}      &  8.83 & 6.25 & 0.46 \\
\bottomrule
\end{tabular}
\end{table}

Three observations align Table~\ref{tab:cost-summary} with Table~\ref{tab:process-stats}.

\textbf{(i) Effort scales with difficulty.}
Easy benchmarks have median wall-clock $\le 8$ minutes and effectively zero Meta-Strategist invocations per problem ($\le 0.13$). Apex 2025 and IMO 2025 invoke the Meta-Strategist $1.6$--$2.2$ times per problem on average, consistent with the high trace-back / re-plan rates in Table~\ref{tab:process-stats}.

\textbf{(ii) Median is much smaller than mean on hard benchmarks.}
On Apex 2025, the mean is $86.7$ minutes but the median is only $35.8$; on IMO 2025, $54.5$ vs.\ $18.6$. A small number of tail problems consume disproportionate time --- in particular, \texttt{apex2025\_12} (which is the same problem as \texttt{imo2025\_6}; see Appendix~\ref{app:case-imop6}) is the longest-running problem in the suite at $498$ minutes ($\approx 8.3$\,h).

\textbf{(iii) Verifier-to-Reasoner call ratio is roughly stable.}
Across all benchmarks the Verifier is invoked at $\approx 0.6$--$0.7\times$ the Reasoner's call count, reflecting the design in Section~\ref{sec:2-3}: each step incurs one Reasoner call followed by Verifier evaluation, plus additional Reasoner calls inside any challenge loop. The Meta-Strategist call count is the most diagnostic of difficulty, varying by two orders of magnitude across benchmarks ($0.02$ on AIME 2025 to $2.16$ on Apex 2025).

\subsection{Per-problem wall-clock distribution}\label{app:stats:distribution}

\begin{figure}[ht]
\centering
\includegraphics[width=0.92\linewidth]{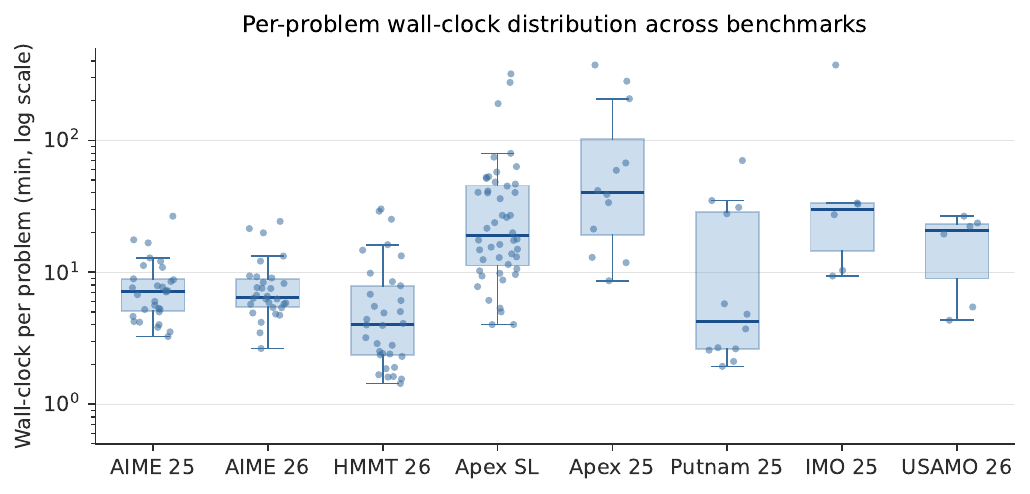}
\caption{Per-problem wall-clock distribution across the eight benchmarks, on a log scale. Each marker is a single problem averaged over its four runs; box-and-whisker overlays show the median and inter-quartile range. AIME, HMMT, and USAMO problems concentrate at $\le 30$\,min; Apex Shortlist, Apex 2025, and IMO 2025 have heavy upper tails dominated by a few hard problems, including the shared $\texttt{apex2025\_12}/\texttt{imo2025\_6}$ tiling problem at the top.}
\label{fig:wallclock}
\end{figure}

Figure~\ref{fig:wallclock} shows the per-problem wall-clock distribution for each benchmark. Each marker is one problem, averaged over its four runs.

The figure makes the head--tail structure visible without relying on standard-deviation summaries (which are unstable at the four-run-per-problem sample size we use).

\subsection{Verification-tag statistics}\label{app:stats:tags}

\begin{figure}[ht]
\centering
\includegraphics[width=0.92\linewidth]{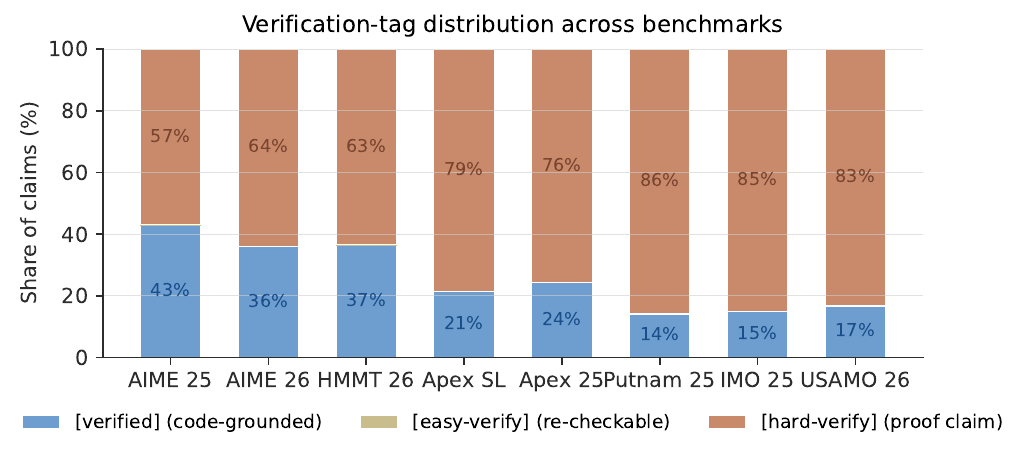}
\caption{Distribution of verification tags across benchmarks, normalized to $100\%$ of claims per benchmark. Final-answer benchmarks (left) carry a substantially larger share of code-grounded \texttt{[verified]} claims; proof-based benchmarks (right) shift toward \texttt{[hard-verify]} claims that require mathematical review.}
\label{fig:tagstack}
\end{figure}

Table~\ref{tab:tag-stats} reports the share of \texttt{[verified]}, \texttt{[easy-verify]}, and \texttt{[hard-verify]} tags across all step reports in the runs of Table~\ref{tab:default-baselines}, and Figure~\ref{fig:tagstack} visualizes the corresponding stacked breakdown.

\begin{table}[ht]
\caption{Verification-tag statistics aggregated over all step reports per benchmark. Counts are over individual tagged claims; percentages are within-benchmark.}
\label{tab:tag-stats}
\centering
\small
\setlength{\tabcolsep}{6pt}
\renewcommand{\arraystretch}{1.05}
\begin{tabular}{@{}lrrrrr@{}}
\toprule
\textbf{Benchmark} & \textbf{\# step reports} & \textbf{\# claims} & \textbf{\% [verified]} & \textbf{\% [easy-verify]} & \textbf{\% [hard-verify]} \\
\midrule
AIME 2025      & 437 &  4{,}572 & 43.0 & 0.07 & 56.9 \\
AIME 2026      & 437 &  4{,}801 & 36.0 & 0.02 & 64.0 \\
HMMT Feb 2026  & 160 &  1{,}988 & 36.5 & 0.05 & 63.4 \\
Apex Shortlist & 682 & 10{,}727 & 21.5 & 0.00 & 78.5 \\
Apex 2025      & 220 &  3{,}329 & 24.2 & 0.00 & 75.8 \\
Putnam 2025    & 117 &  1{,}930 & 14.1 & 0.00 & 85.9 \\
IMO 2025       & 106 &  2{,}088 & 14.8 & 0.00 & 85.2 \\
USAMO 2026     & 108 &  1{,}860 & 16.7 & 0.00 & 83.3 \\
\bottomrule
\end{tabular}
\end{table}

The pattern is consistent with Section~\ref{sec:2-3}'s claim that the verification-tag protocol implements a principled trade-off between reasoning and tools rather than a uniform bias toward either. \texttt{[verified]} claims dominate on AIME 2025/2026 ($36$--$43\%$) and HMMT 2026 ($36.5\%$), where final-answer arithmetic and small-case enumeration can be discharged by code; on the proof-based benchmarks (Putnam 2025, IMO 2025, USAMO 2026) the share of \texttt{[hard-verify]} claims rises to $83$--$86\%$, which is precisely where Section~\ref{sec:2-3}'s Logic Gate carries the bulk of verification work. \texttt{[easy-verify]} is rare ($\le 0.07\%$) across the board: when the Reasoner can cheaply check a claim, it almost always runs the check itself and upgrades the tag to \texttt{[verified]}, the behaviour explicitly mandated by the Reasoner system prompt (Appendix~\ref{app:reasoner}). This turns the verification-tag protocol from a design choice into a measured property: the share of code-grounded claims tracks problem type, and the Verifier's calibration is exercised on a measurable mix of evidence.

\section{Extended Case Study I: Apex 2025 Problem 2}\label{app:case-apex2}

This appendix complements Section~\ref{sec:case-apex2} by laying out STAR-PólyaMath's full trajectory for Apex 2025 Problem~2. We trace the system's evolution across three phases: Plan~v1 (a failed attempt anchored on an incorrect $3/4$ bound), the persistent Meta-Strategist's diagnoses across three trace-backs and the final re-plan verdict, and Plan~v2 (the successful re-execution targeting the correct $1/2$ constant).

\subsection{Plan~v1: the failed $3/4$ attempt}\label{app:apex2:v1}

Plan~v1 conjectured $k = 3/4$ and committed to proving the universal bound $a_1 + a_2 > \tfrac{3}{4}\, a_3$. The evidence came from the \emph{comb family} of polyominoes (a horizontal spine of $2m+1$ unit squares with one square attached above and below every other interior spine square). The verified comb counts, recorded in \texttt{PROBLEM\_STATE.md} under ``Rescued from earlier attempts (Plan v1, Step~3)'', are
\[
a_1(C_m)=2m+2,\; a_2(C_m)=m+3,\; a_3(C_m)=4m,
\;
\frac{a_1(C_m)+a_2(C_m)}{a_3(C_m)}=\frac{3}{4}+\frac{5}{4m}\xrightarrow{m\to\infty}\frac{3}{4}.
\]
The Reasoner correctly established these counts (verified by a Step~3 enumeration script) and confirmed by exhaustive search that no hole-free fixed polyomino through $10$ cells violates $4(a_1+a_2)\le 3a_3$. The candidate constant $3/4$ was therefore strongly supported by the evidence available at that point.

Step~4's objective --- to prove $a_1+a_2>\tfrac{3}{4}a_3$ universally --- repeatedly timed out. Each Reasoner attempt set up a local discharging argument over boundary-adjacent square types but ran into unbounded propagation involving two-edge squares with opposite boundary edges. Three successive Step-4 attempts ended in timeouts, each triggering a Verifier-then-Meta-Strategist trace-back. The first trace-back diagnosis read:

\begin{quote}\small\itshape
Trace-back to Step~4: Prove the universal lower bound. The timeout is not a computation problem; the Reasoner's partial report shows the current \textbf{purely local discharging plan is structurally insufficient} because the double-supported $a_2$ obstruction can propagate. The next attempt should stop trying to close the same local table and instead model the obstruction chains globally.
\end{quote}

\subsection{Three trace-backs and the re-plan verdict}\label{app:apex2:meta}

All four interventions below originated from the \emph{same} persistent Meta-Strategist session. This is the mechanism the paper highlights in Section~\ref{sec:case-apex2}: a non-persistent supervisor that sees only one timeout at a time would treat each as a local proof gap, but evidence accumulated within a single session reveals a structural pattern.

\paragraph{First trace-back (Step~4).}
\begin{quote}\small\itshape
The timeout is not a computation problem; the current purely local discharging plan is structurally insufficient because the double-supported $a_2$ obstruction can propagate. \textsc{Action\_Type:} \textsc{Trace\_Back}; \textsc{Trace\_Back\_To:} 4; \textsc{Use\_Pure\_Reasoning:} YES.
\end{quote}

\paragraph{Second trace-back (Step~4).}
\begin{quote}\small\itshape
The failure is strategic, not computational: the local discharging table keeps running into the same unbounded strip-propagation obstruction. The Reasoner should stop extending the local table and convert the obstruction into a component-level accounting problem. \textsc{Action\_Type:} \textsc{Trace\_Back}; \textsc{Trace\_Back\_To:} 4; \textsc{Use\_Pure\_Reasoning:} YES.
\end{quote}

\paragraph{Third trace-back (Step~4).}
\begin{quote}\small\itshape
The failure is localized to Step~4's execution method: repeated timeouts indicate the current computational attack is infeasible within the step budget, but there is not enough evidence that the earlier plan structure or prior verified steps are unsound. Re-execute Step~4 from scratch with a revised, more structural or targeted approach rather than approving a full re-plan.
\end{quote}

After the third trace-back also yielded a timeout, the Meta-Strategist promoted the diagnosis from local-repair to plan-level: the bound itself was false, not unproved. The re-plan verdict reads:

\begin{quote}\small\itshape
\textbf{Plan summary.} Plan vK pursued a strategy anchored on proving a universal $3/4$ lower bound, likely treating that value as the candidate sharp threshold or final answer. This fails because the proposed universal bound is false, so further patching of the same inequality direction would continue to reinforce an invalid target.

\textbf{Reason.} The current plan is unsound because its answer-bearing target was to establish a universal lower bound of $3/4$, and the Reasoner now reports that this bound is false rather than merely unproved at Step~4. This is not a local proof gap suitable for re-executing one step; the target itself must be replaced.
\end{quote}

The verdict carried explicit \emph{forbidden directions} (verbatim):

\begin{lstlisting}[basicstyle=\ttfamily\footnotesize,breaklines=true,frame=single]
- Do not assume 3/4 is the answer or the correct universal lower bound.
- Do not start the next plan by trying to repair the proof of the
  3/4 lower bound.
- Do not prove any inequality whose main target is a universal RHS of
  3/4 before independently re-establishing the correct candidate value.
- Do not reuse lemmas whose only role is to force or certify the false
  3/4 bound.
- Do not restrict construction search to cases consistent with the 3/4
  target; the next plan must actively look for a replacement target
  from fresh evidence.
\end{lstlisting}

These directives are injected into every subsequent Reasoner spawn through the \texttt{PROBLEM\_STATE.md} hook (Appendix~\ref{app:impl:hooks}), so v2 cannot silently re-anchor on $3/4$.

\subsection{Plan~v2: the successful $1/2$ trajectory}\label{app:apex2:v2}

Plan~v2's seven steps are:

\begin{enumerate}\setlength{\itemsep}{0pt}
\item Fix the exact combinatorial model.
\item Split the counts into inside and outside contributions.
\item Find the correct candidate constant from fresh evidence.
\item Prove the universal lower bound.
\item Make the discharging proof independent of shape pathologies.
\item Construct polygons approaching equality.
\item Conclude sharpness and the largest value.
\end{enumerate}

\paragraph{Step~3 (new candidate).} Re-running candidate search on a denser $4{\times}4$ motif (rows \texttt{\#.\#.}, \texttt{\#\#\#\#}, \texttt{\#.\#.}, \texttt{....}) tiled $n\times n$ produced
\[
a_1=9n,\quad a_2=4n^2-n,\quad a_3=8n^2-n,
\]
and for $n=50$ the ratio is $208/399\approx 0.5213$ --- already well below $3/4$ at $n=5$.

\paragraph{Step~4 (cap-map proof).} The verified lemmas establishing $a_3\le a_1+2a_2$ are recorded in the Verified Results ledger:

\begin{quote}\small\itshape
[Lemma] Around any lattice vertex, the four adjacent squares cannot alternate inside-outside-inside-outside cyclically.

[Lemma] If $e(Q)=3$ and $f(Q)$ is the square across the boundary side opposite the unique same-colored neighbor of $Q$, then $e(f(Q))\in\{1,2\}$.

[Lemma] For any square $R$ with $e(R)\in\{1,2\}$, at most $e(R)$ three-edge squares map to $R$ under the cap-target map.

[Lemma] Consequently, $a_3 \le a_1 + 2 a_2$.
\end{quote}

The cap-map inequality was independently \texttt{[verified]} by a Step~4 verification script, which reported \emph{zero failures over $6{,}234$ generated simply connected polyominoes up to size $10$}, plus all listed motif cases.

\paragraph{Step~5 (pathology audit).} A Step~5 audit script confirmed that the cap-map argument is independent of shape pathologies (interior ears, exterior notches, opposite-edge two-squares), reporting zero bad cap targets and zero capacity failures across the audit.

\paragraph{Step~6 (sharpness construction).} A bridged $4{\times}4$-motif family $\{S_n\}$ gives admissible simple lattice polygons with
\[
(a_1,a_2,a_3)=(11n-2,\ 4n^2+n-2,\ 8n^2-3n+2),
\quad
\frac{a_1+a_2}{a_3}=\frac{1}{2}+\frac{27n-10}{16n^2-6n+4}\xrightarrow{n\to\infty}\frac{1}{2}^{+}.
\]
An independent verifier script confirmed the formula and validity for $n=1,\ldots,40$.

\paragraph{Step~7 (conclusion).} Combining Step~4 (every admissible polygon satisfies $a_1+a_2>\tfrac{1}{2}a_3$) and Step~6 (a family realizing the ratio $\to 1/2$), no $k>1/2$ works and $k=1/2$ does. Plan~v2 was accepted at Step~7 without further intervention.

\subsection{Final solution}\label{app:apex2:sol}

The accepted answer is $\boxed{k=\dfrac{1}{2}}$. The cap-map argument from the final solution reads:

\begin{quote}\small
Color squares inside $P$ black and squares outside $P$ white; a boundary edge separates unlike colors. Around any lattice vertex the four squares cannot alternate black-white-black-white (else four boundary edges meet there).

Take any square $Q$ with $e(Q)=3$; it has a unique same-colored neighbor. Let $f(Q)$ be the square across the boundary edge \emph{opposite} that neighbor. By the alternation impossibility, $e(f(Q))\in\{1,2\}$. For any target $R$ with $e(R)\in\{1,2\}$, at most $e(R)$ three-edge squares cap onto $R$, so
\[
a_3 \le a_1 + 2a_2.
\]
The lowest horizontal boundary edge of $P$ certifies $a_1>0$; hence
\[
a_3 \le a_1+2a_2 < 2(a_1+a_2),
\quad\text{i.e.,}\quad a_1+a_2 > \tfrac{1}{2}\, a_3.
\]
Combined with the $4{\times}4$ motif construction approaching ratio $1/2$ from above, the largest $k$ is $1/2$.
\end{quote}

Full step reports, debate transcripts, and verification scripts are archived in the supplementary code release.

\section{Extended Case Study II: Apex 2025 Problem 12 (IMO 2025 Problem 6)}\label{app:case-imop6}

Apex 2025 Problem~12 is identical to IMO 2025 Problem~6, the $2025{\times}2025$ tiling problem. It is the hardest combinatorics problem in our benchmark suite. STAR-PólyaMath solved it end-to-end in a single run: the accepted solution arrived in plan version 3, with $8$ steps, $12$ cumulative trace-backs, $2$ re-plans, and a total wall-clock time of $8$h $18$m.

\subsection{Problem and answer}\label{app:imop6:problem}

\begin{quote}\small
Consider a $2025\times 2025$ grid of unit squares. Matilda wishes to place on the grid some rectangular tiles, possibly of different sizes, such that each side of every tile lies on a grid line and every unit square is covered by at most one tile. Determine the minimum number of tiles Matilda needs to place so that each row and each column of the grid has exactly one unit square that is not covered by any tile.
\end{quote}

The accepted answer is $\boxed{2112}$.

\subsection{Plan and trajectory}\label{app:imop6:plan}

The accepted plan~v3 has eight steps:

\begin{enumerate}\setlength{\itemsep}{0pt}
\item Recast the grid condition exactly.
\item Replace tilings by a rectilinear-geometry invariant.
\item Specialize good chords to permutation geometry.
\item Translate chord selection into a bipartite matching problem.
\item Independently search for the extremal construction.
\item Prove the construction's upper bound geometrically.
\item Prove the matching lower bound for every permutation.
\item Assemble the final equality.
\end{enumerate}

\paragraph{Reformulation phase (Steps~1--3).} Step~1 establishes that any valid configuration has uncovered set $U_\pi=\{(i,\pi(i)):1\le i\le 2025\}$ for some permutation $\pi\in S_{2025}$, and conversely; the problem reduces to $\min_\pi T(\pi)$, where $T(\pi)$ is the minimum number of rectangles partitioning the complement of $U_\pi$.

Step~2 entered a multi-round debate. The Verifier challenged an initial formula $T(\pi)=r(\pi)/2+h(\pi)+1-\nu(\pi)$, citing a small-case counterexample (the $n=3$ identity permutation gave a predicted $3$ tiles vs.\ a true count of $4$). The Reasoner revised the formula to
\[
T(\pi)=r(\pi)+c(\pi)-h(\pi)-\nu(\pi),
\]
where $r$ counts resolved reflex corners, $c$ counts covered connected components, $h$ counts interior holes, and $\nu(\pi)$ is the maximum number of pairwise non-crossing \emph{good chords}. The corrected formula was accepted after the debate.

Step~3 specializes good chords to permutation geometry: a horizontal chord exists between rows $i$ and $i+1$ exactly when $|\pi(i+1)-\pi(i)|\ge 2$, with an analogous criterion for vertical chords on adjacent columns. Compatible (pairwise non-crossing) chords form an independent set in a bipartite incompatibility graph $G_\pi$.

\paragraph{Reduction phase (Steps~4--5).} Step~4 applies K\"onig's theorem to the bipartite graph: $\nu(\pi)=|\mathcal{H}|+|\mathcal{V}|-\mu(G_\pi)$ where $\mu(G_\pi)$ is the maximum matching. Combining with Step~2 yields
\[
T(\pi) = n + a(\pi) + \mu(G_\pi) - 1,
\]
where $a(\pi)$ counts diagonally adjacent pairs of consecutive uncovered cells and $n=2025$.

Step~5 entered a second multi-round debate. The Verifier rejected an initial bare existence claim of a $2112$-tile construction and demanded an explicit permutation rule, computed value, and verification script outputs. The Reasoner responded with the construction: take $m=45$ (so $2025=m^2$) and define
\[
\pi(am+i+1)=im+(45-a),\qquad 0\le a,i<m.
\]
This permutation, together with the four boundary rectangle families and $(m-1)^2$ central rectangles, partitions the complement into exactly $4(m-1)+(m-1)^2=2112$ rectangles. A Step~5 verification script confirmed coverage and disjointness.

\paragraph{Lower-bound phase (Steps~6--7).} Step~6 enumerates the $2112$ rectangles explicitly, with a Step~6 verification script reporting \texttt{rectangles=2112}, \texttt{covered=4098600}, \texttt{expected\_covered=4098600}, hence $\min_\pi T(\pi)\le 2112$.

Step~7 proves the matching lower bound for \emph{every} permutation. Let $X\cup Y$ be a minimum vertex cover of $G_\pi$ with $|X|=x$, $|Y|=y$; the $x$ row cuts and $y$ column cuts partition the grid into $(x+1)(y+1)$ zones. Each diagonal-adjacency component lies within a single zone (diagonal adjacencies admit no good chord), so $n-a(\pi)\le (x+1)(y+1)$, which algebraically forces
\[
a(\pi)+x+y\ge 2\sqrt{n}-2.
\]
For $n=2025=45^2$ this gives $a(\pi)+\mu(G_\pi)\ge 88$, hence
\[
T(\pi)=n+a(\pi)+\mu(G_\pi)-1\ge 2025+88-1=2112.
\]

Step~8 assembles the equality $\min_\pi T(\pi)=2112$.

\paragraph{Re-plan history.} The two re-plans recorded in \texttt{PROBLEM\_STATE.md} both abandoned earlier targets that were directly falsified by computational evidence. The most consequential, recorded under ``Confirmed Failures'', reads:

\begin{quote}\small\itshape
Plan vK pursued a proof of a $2700$ lower bound for the tiling quantity; this fails because Step~4 produced a verified $2112$-tile construction, giving a concrete counterexample to the intended lower bound and invalidating the plan's main direction. The current plan is unsound because its planned $2700$ lower bound has been directly falsified by a verified $2112$-tile construction. This is not a step-local repair issue: the central target inequality of the plan is false, so tracing back within the same plan would keep the Reasoner anchored to an impossible bound.
\end{quote}

The accompanying forbidden direction (``Do not attempt to prove a $2700$ lower bound or any lower bound exceeding the verified $2112$-tile construction.'') was injected into the next plan, ensuring the system did not re-anchor on the falsified target.

\subsection{Key verified results}\label{app:imop6:verified}

The Verified Results ledger records the following representative entries (verbatim from \texttt{PROBLEM\_STATE.md}):

\begin{itemize}[leftmargin=*,itemsep=0pt,topsep=2pt]
  \item \textbf{[Lemma]} (Step~1) Valid configurations are in bijection with permutations $\pi\in S_{2025}$ via $U_\pi=\{(i,\pi(i))\}$.
  \item \textbf{[Lemma]} (Step~2) $T(\pi)=r(\pi)+c(\pi)-h(\pi)-\nu(\pi)$.
  \item \textbf{[Computation]} (Step~3) A Step~3 verification script confirms the chord classification for all permutations of size $n=2,\ldots,6$.
  \item \textbf{[Computation]} (Step~4) A Step~4 verification script confirms the bipartite matching formula for $n=2,\ldots,7$.
  \item \textbf{[Computation]} (Step~5) The number of rectangles is $4(m-1)+(m-1)^2=2112$ for $m=45$.
  \item \textbf{[Computation]} (Step~6) A Step~6 verification script confirms \texttt{rectangles=2112}, \texttt{covered=4098600}, \texttt{expected\_covered=4098600}.
  \item \textbf{[Lemma]} (Step~7) For every $\pi\in S_{2025}$, $T(\pi)\ge 2112$, established by three independent Step~7 verification scripts (exact formula on small cases, vertex-cover slicing, matching bound).
  \item \textbf{[Answer]} (Step~8) The minimum number of rectangular tiles Matilda needs is $2112$.
\end{itemize}

\subsection{Final solution sketch}\label{app:imop6:sol}

\begin{quote}\small
\textbf{Upper bound (construction).} With $m=45$ and $\pi(am+i+1)=im+(45-a)$, the complement of $U_\pi$ is partitioned by four boundary rectangle families plus $(m-1)^2$ central rectangles, totaling $4(m-1)+(m-1)^2=2112$ rectangles. Hence $\min_\pi T(\pi)\le 2112$.

\textbf{Lower bound (matching argument).} For any $\pi\in S_{2025}$, $T(\pi)=n+a(\pi)+\mu(G_\pi)-1$ where $a(\pi)$ counts diagonal adjacencies of consecutive uncovered cells and $\mu(G_\pi)$ is the maximum matching of the bipartite incompatibility graph of good chords. A vertex-cover slicing argument shows $a(\pi)+\mu(G_\pi)\ge 2\sqrt{n}-2$. For $n=2025=45^2$, this gives $a(\pi)+\mu(G_\pi)\ge 88$, so $T(\pi)\ge 2025+88-1=2112$ for every $\pi$.

\textbf{Conclusion.} Upper and lower bounds coincide, so $\min_\pi T(\pi)=2112$.
\end{quote}

The complete proof, all step reports, debate transcripts, and verification scripts are archived in the supplementary code release.


\end{document}